\documentclass[structabstract]{aa}  

\usepackage{graphicx}
\usepackage{txfonts}
 \usepackage{amssymb}
 \usepackage{natbib}
  \usepackage{lscape}
  \bibpunct{(}{)}{;}{a}{,}{,}
\usepackage{hyperref}
  \usepackage{times}

  \def\deg{${}^\circ$}
  \def\min{${}^{\prime}$}
  
 \def\memrv{893 }
 \def\memall{661 }
 \def\memprob{9 }
 \def\memconf{652 }
 \def\memctts{333 } 
 \def\memwtts{328 }
 \def\memcttsage{240 } % diagrams_pisa
 \def\memwttsage{147 } % diagrams_pisa
  \def\giga{237}
  \def\gigb{33}
  \def\gigc{11}
  \def\gigtot{281 }
  \def\gigrc{117 }
  \def\gignorc{164 }
  \def\teffwfi{395 }
  \def\teffvphas{364}
  \def\teffwfivphas{337 }
  \def\clustdist{1320} % see cluster_distance.pro
  \usepackage{longtable}
\usepackage{longtable,lscape}
  \usepackage{lscape}
\def\hii{\ion{H}{ii} }
\begin{document}
   \title{The Gaia-ESO Survey:  age spread in the star forming 
region NGC\,6530 from the HR diagram and gravity indicators
 \thanks{Based on
   observations made with the ESO/VLT, at Paranal Observatory, 
   under program 188.B-3002 (The Gaia-ESO Public Spectroscopic Survey)  }}

   \subtitle{ }

  \author{L. Prisinzano\inst{\ref{oapa}}
          \and
          F. Damiani\inst{\ref{oapa}}
          \and
          V. Kalari\inst{\ref{unicile}, \ref{armagh}}
           \and
          R. Jeffries\inst{\ref{keele}}
         \and
          R. Bonito\inst{\ref{oapa}}
          \and
          G. Micela\inst{\ref{oapa}}
           \and
          N. J. Wright\inst{\ref{keele}}
          \and
          R. J. Jackson\inst{\ref{keele}}  
          \and       
          E. Tognelli\inst{\ref{pisa}}
           \and
          M. G. Guarcello\inst{\ref{oapa}}
         \and
          J. S. Vink\inst{\ref{armagh}}
	    \and
	    A. Klutsch\inst{\ref{oact}}
	            \and
            F. M. Jim\'enez-Esteban\inst{\ref{esac}}
          \and       
          V. Roccatagliata\inst{\ref{pisa}, \ref{arcetri}, \ref{infn}}
           \and
            G. Tautvai\v{s}ien\.{e}\inst{\ref{vilnius}}
           \and
          G. Gilmore\inst{\ref{cambridge}}
	    \and
	    S. Randich\inst{\ref{arcetri}}
		 \and
         E. J. Alfaro\inst{\ref{granada}}         %
         \and
          E. Flaccomio\inst{\ref{oapa}}
 	\and
	 S. Koposov\inst{\ref{cambridge}}
         \and
	   A. Lanzafame\inst{\ref{unict}}
	  \and
          E. Pancino\inst{\ref{arcetri}}
	  	 \and
	 M. Bergemann\inst{\ref{maxplank}}
	 \and 	 
	  G. Carraro\inst{\ref{unipd}}
        \and 
         E. Franciosini\inst{\ref{arcetri}}
         \and
          A. Frasca\inst{\ref{oact}}
	\and
	  A. Gonneau\inst{\ref{cambridge}}
	  \and 
	  A. Hourihane\inst{\ref{cambridge}}
           \and
           P. Jofr\'e\inst{\ref{ciencias}} 
         \and
	 J. Lewis\inst{\ref{cambridge}}
          \and
           L. Magrini\inst{\ref{arcetri}}
          \and 
          L. Monaco\inst{\ref{ciencias}}
	  \and 
	  L. Morbidelli\inst{\ref{arcetri}}
 	  \and
        G. G. Sacco\inst{\ref{arcetri}}
	 \and 
	 C.C. Worley\inst{\ref{cambridge}}
	 \and
           S. Zaggia\inst{\ref{oapd}}           }
   \institute{INAF - Osservatorio Astronomico di Palermo, Piazza del Parlamento 1, 90134, Palermo, Italy \\ %1
              \email{loredana.prisinzano@inaf.it}\label{oapa}
       \and
       Departamento de Astronomía, Universidad de Chile, Casilla 36-D, Santiago, Chile\label{unicile}
	\and
        Armagh Observatory and Planetarium, College Hill, Armagh, BT61 9DG, UK\label{armagh}
       \and
        Astrophysics Group, Keele University, Keele, Staffordshire ST5 5BG, United Kingdom\label{keele} %2
        \and
        Department of Physics `E. Fermi', University of Pisa, Largo Bruno Pontecorvo 3, I-56127 Pisa, Italy\label{pisa}
        \and
        INAF - Osservatorio Astrofisico di Catania, via S. Sofia 78, 95123, Catania, Italy\label{oact} %4
	\and
	Departmento de Astrof\'{\i}sica, Centro de Astrobiolog\'{\i}a (INTA-CSIC), ESAC Campus, Camino Bajo del 
	Castillo s/n, E-28692 Villanueva de la Ca\~nada, Madrid, Spain\label{esac}
        \and
        INAF - Osservatorio Astrofisico di Arcetri, Largo E. Fermi 5, 50125, Florence, Italy\label{arcetri} %3
	    Florence, Italy
         \and
	 INFN, Sezione di Pisa, Largo Pontecorvo 3, 56127 Pisa, Italy\label{infn}	
	\and
	Astronomical Observatory, Institute of Theoretical Physics and Astronomy, Vilnius University, 
	Saul\.{e}tekio av. 3, 10257 Vilnius, Lithuania\label{vilnius}
         \and
	 Institute of Astronomy, University of Cambridge, Madingley Road, Cambridge CB3 0HA, United Kingdom\label{cambridge} %11
 	\and
        Instituto de Astrof\'{i}sica de Andaluc\'{i}a-CSIC, Apdo. 3004, 18080 Granada, Spain\label{granada} %6
        \and
        Dipartimento di Fisica e Astronomia, Universit\`a di Catania, via S. Sofia 78, 95123, Catania, Italy\label{unict} %5 
        \and
	Max Planck Institute for Astronomy, Koenigstuhl 17, 69117 Heidelberg, Germany\label{maxplank}
	\and
        Dipartimento di Fisica e Astronomia Galileo Galilei, Universit\`a di Padova, Vicolo Osservatorio 3, 
        I-35122, Padova, Italy\label{unipd}
	\and
	Departamento de Ciencias Fisicas, Universidad Andres Bello, Republica 220, Santiago, Chile\label{ciencias} %13
	    \and 
 	INAF - Padova Observatory, Vicolo dell'Osservatorio 5, 35122 Padova, Italy\label{oapd}    %9
%     	    \and 
%         Astrophysics Research Institute, Liverpool John Moores University, 146 Brownlow Hill, Liverpool L3 5RF, 
%         United Kingdom\label{live} %12
%	\and 
%	Department of Astronomy, University of Geneva, 51 chemin des Maillettes, 1290, Versoix, Switzerland\label{geneva} %14
         }
 
    %\date{May 2015}
% \abstract{}{}{}{}{} 
% 5 {} token are mandatory
 
  \abstract
{In very young clusters, the stellar age distribution is the empirical proof
of the duration of star cluster formation and thus it gives indications of
the physical mechanisms involved in the star formation process. 
Determining  the amount of interstellar extinction and the
correct reddening law
are crucial steps to derive  fundamental stellar parameters and in particular accurate
ages from the HR diagram. }  
  {In this context, we  derived accurate stellar ages 
for  NGC\,6530, the young cluster associated with the Lagoon Nebula
to infer the star formation history of this region.}
 {We use the \emph{Gaia}-ESO survey observations of the Lagoon Nebula,
together with photometric literature data and \emph{Gaia} DR2 kinematics, 
to derive cluster membership and fundamental stellar parameters.
Using spectroscopic effective temperatures,  
we analyze the reddening properties of all objects  and derive accurate
stellar ages for cluster members.}
   {We identified \memconf  confirmed and \memprob probable members.
The reddening inferred for members and non-members allows us to distinguish
foreground objects, mainly main-sequence (MS) stars,  and background objects, mainly giants.
This classification is  in agreement with the distances inferred from 
\emph{Gaia} DR2 parallaxes for these objects. The foreground and background stars
show a spatial pattern that allows us to trace the  three-dimensional structure
of the nebular dust component. 
Finally,
we derive stellar ages for 382 confirmed cluster members
for which we obtained the individual reddening values. 
In addition, we find that the gravity-sensitive $\gamma$ index distribution
for the M-type stars is correlated with stellar age.
}
   {For all members with T$_{\rm eff}<5500$\,K, the mean logarithmic  age  
   is 5.84 (units of years) with a dispersion of 0.36 dex.
   The age distribution of stars with accretion and/or disk (CTTSe) is similar 
   to that of stars without accretion and without disk (WTTSp). 
 We interpret this dispersion as evidence of a real 
 age spread since the total uncertainties on age determinations,
  derived from Monte Carlo simulations,
 are significantly smaller than the observed spread. This conclusion 
  is supported by the evidence of a decreasing of the gravity-sensitive $\gamma$ index
 as a function of stellar ages. The presence of a small age spread is
also supported by
 the spatial distribution  and the kinematics of old and young members. 
In particular, members with accretion and/or disk, 
 formed in the last 1\,Myr, show evidence
of subclustering around the cluster center, in the Hourglass Nebula
and in the M8-E region, suggesting 
 a possible triggering of star formation
  events by the O-type star ionization fronts.}

   \keywords{stars: pre-main sequence -- stars: stellar evolution --HR diagram-- 
   stars: age spread --reddening --reddening law- open clusters and associations:
    individual: NGC\,6530, stars: formation
               }
\authorrunning{L. Prisinzano et al.}
\titlerunning{Age spread in the star forming region NGC\,6530}
   \maketitle
%
%________________________________________________________________

\section{Introduction}
Optical and infrared (IR) observations obtained in the last two decades clearly show  
that stars form mainly in groups within giant molecular
clouds \citep[e.g.][]{gute09,moli10}. However, it is still under debate if
the process is moderated by turbulence and magnetic fields
 \citep{tan06} or
if it is rapid and efficient, taking place on dynamical timescales
\citep{elme00,elme07}. In the first case,
 clouds can sustain   the production
of stars for a period of at least several dynamical timescales  ($\sim10^7$ yr)
while in the second case, star formation occurs in a free-fall time 
\citep[t$_{\rm ff}\sim10^6$ yr,][]{macl04,tass04}.

 Estimating stellar ages and  age spreads in young clusters is a crucial test to
 understand how star formation occurs over time.
Accurate isochronal ages derived from the HR diagrams of young clusters allow us  
to efficiently reconstruct the star formation history.
A first hint of luminosity spread, associated with an 
age spread, was noted by \citet{pall99} 
 who concluded that the contraction of the parent cloud 
 in the Orion Nebula Cluster   started 10$^7$ yrs
ago. The process
 proceeded gradually  with an accelerating star formation rate, forming the bulk 
of the stars in the last 1-2\,Myr.

A similar luminosity spread has also been observed  in other young clusters but
some authors have explained it
as an effect 
of observational uncertainties \citep{hill98,hart01,sode14},
or related to the oversimplified stellar models adopted.
 In particular, \citet{bara09} 
 showed that the inclusion of protostellar accretion 
 in the computation of stellar models can produce a  
  luminosity spread, which in turn reflects in an apparent 
  age spread up to about 10 Myr \citep[see also][]{bara12}.
 From the observational point of view,
the precision of the stellar ages 
strongly depends on uncertainties in  extinction, non-photospheric effects due to accretion 
 or circumstellar disks, but also 
on uncertainties in distance, intrinsic young star variability and binarity. 
In  young clusters still surrounded by the parent molecular cloud, extinction is expected to
be non-uniform, due to the inhomogeneity of the material around the stars.
 In addition, it has been shown in several cases \citep{walk57,da-r16}, that 
due to grain growth, in star forming regions the non standard reddening law R$_{\rm V}=5$,
 rather than the standard R$_{\rm V}$=3.1,
gives a better description of the interstellar absorption.
The adoption of an unsuitable reddening law can therefore 
introduce an artificial dispersion.

\citet{prei12} performed a simulation of a young cluster and
 concluded that observational uncertainties cannot explain the entire extent
of the spread observed in the HR diagrams  and  in some cases it has been proven that
the observed luminosity spread is due to a real spread of the projected radii,
for stars of a given effective temperature \citep{jeff07}.

A further and independent way to test the reality of the age spread  is the Li depletion in low-mass
stars,  an age-dependent process, that is also function of luminosity and   effective temperature.
The first evidence of age spread based on the lithium test was presented 
by \citet{pall07}. Later, several other works have confirmed the presence of a dispersion of the Li abundance,
even though, the method is model-dependent regarding mass estimates.

In a recent review, \citet{jeff17a} concludes that all the adopted methods 
(HR diagram, Li depletion,
and spectroscopic radii)  suggest that, 
uncertainties alone cannot account for the luminosity spreads that are seen.
That could point to (relatively modest) age spreads or it could be highlighting
 deficiencies in PMS models. 
 Finally, using spectroscopic IR data,
\citet{da-r16} found a correlation between HR-diagram ages and ages inferred from 
spectroscopic gravity indicators that strongly suggest an radius spread 
 in the Orion\,A molecular cloud.
 
Multi-object spectrographs, such as the ESO VLT/FLAMES, used for the
 Gaia-ESO Survey 
\citep[GES--][]{gilm12,rand13}, provide very useful datasets to significantly reduce
the large uncertainties affecting stellar parameters involved compiling an HR diagram.
Such kind of data, associated with other  optical/near-infrared (NIR) photometric and X-ray  data,
are  not only pivotal to assert cluster membership, 
but allow us to  derive effective temperatures.
This  is a fundamental step to derive individual stellar extinctions
 and then to accurately place the objects in the
HR diagram and compare them to theoretical models. 

We present in this work an  analysis based on the
 Gaia-ESO survey spectroscopic data assembled together with available literature
data   of NGC\,6530, the young cluster associated with the 
\hii region known as Lagoon Nebula.
Evidence of ongoing star formation has been found in this region
 not only around the cluster
center but also in the Hourglass Nebula and the M8 E region, 
through X-ray, H$\alpha$ emission, NIR excesses,
and submillimetre-wave emission, as reviewed by \citet{toth08}. 
The region includes several O-type stars,
associated with the young cluster NGC\,6530, located at about 1250\,pc from the
Sun \citep{pris05}, that is superimposed on the Eastern half 
of the \hii region.   
The region is characterized by a dark lane, splitting the optical nebula, and several bright rims.

Deep X-ray and optical observations allowed us to discover a very large
 population of low mass members
 \citep{sung00,dami04,pris05,dami06} with evidence of a sequential 
 star formation. However, such
evidence was not  found by  \citet{kala15}, based on deep 
VPHAS+  optical data \citep{drew14}
and focused on stars showing H$\alpha$ excesses.
%leaving open the issue of the star formation history in the region.     
A still  debated question, 
strongly related to the determination of stellar ages in this cluster,
 is the reddening law 
holding for the young objects of NGC\,6530. Several studies adopted 
an anomalous reddening that is R$_{\rm V}=$5.0 \citep[e.g.][]{mcca90,kuma04}
rather than the canonical value R$_{\rm V}$=3.1, that is generally valid for the
 Galactic Plane.
The larger value is expected to be more appropriate for star-forming regions,
 since larger dust grains
are not as efficient at blocking blue light as smaller grains towards 
more evolved clusters.

In this work, we want to constrain the reddening law towards NGC\,6530
with the aim of deriving accurate stellar ages, by exploiting
  a very large and unbiased sample of   cluster members 
  observed spectroscopically.

In Sect.\,2 we present the observational data, 
while in Sect.\,3 we describe how we assembled the adopted data. 
 In Sect.\,4
we present the membership criteria used 
to define the final list of confirmed members, while in Sect.\,5 we classify the
remaining Gaia-ESO survey targets as giants or MS stars. 
In Sect.\,6 we analyse the interstellar reddening affecting cluster members 
and contaminants from which we derive hints on the 
structure and thickness of the Lagoon Nebula and in Sect.\,7 we present spectroscopic and
photometric evidences in favor of the anomalous (standard) 
 reddening law for cluster members (background giants). In Sect.\,8 we analyse
the  spatial distributions of cluster members and foreground and background contaminants
in the context of the Lagoon Nebula while in Sect.\,9 we present 
stellar ages for the cluster members and their correlation with the 
gravity-sensitive $\gamma$ index.
Finally, a discussion of our results is presented in Sect.\,10, while
 our concluding remarks are summarized  in Sect.\,11.

%__________________________________________________________________
\section{Observational data}
\begin{figure*}
 \centering
 \includegraphics[width=\textwidth]{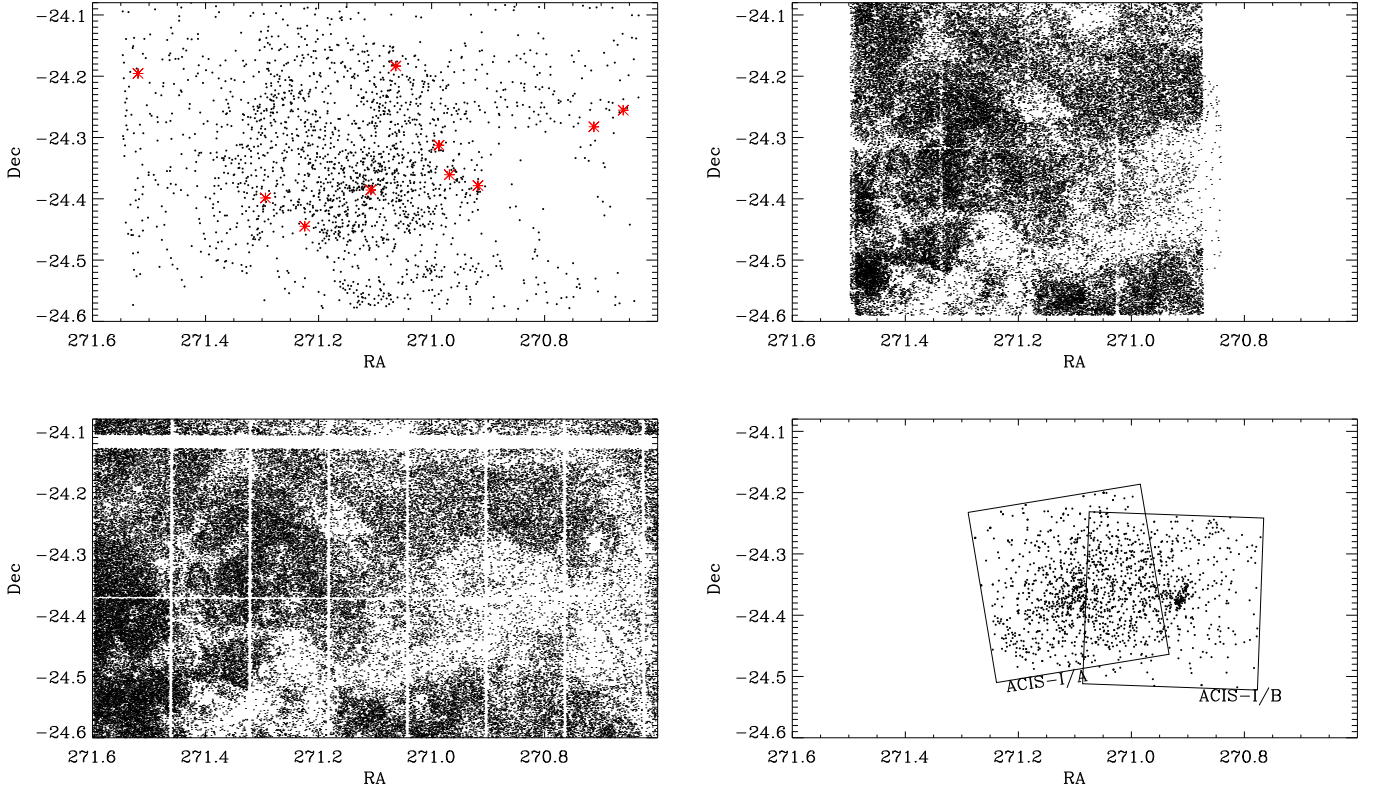} 
 \caption{Spatial distribution of GES targets (black dots) and OB stars (red symbols) 
(upper/left panel), WFI and VPHAS+ sources (upper/right and 
lower/left panels, respectively) and X-ray detections (lower/right panel).}
\label{plotacis}
 \end{figure*}

The dataset used in this work includes  the 
GES spectroscopic data, several  optical and NIR photometric catalogs
from the literature and a new list of Chandra X-ray detections.
Details are given in the following sections.

\subsection{Gaia-ESO survey spectroscopic  data}
Spectroscopic data used in this work were acquired within  GES,
using simultaneously the instruments GIRAFFE and UVES of the ESO VLT/FLAMES
 multi-fibre spectrograph 
\citep{pasq02}. 
The observations of the NGC\,6530 region were carried out during 17 nights, in September 2012
and June-September 2013, using the setups listed in Table\,\ref{setuptab}. To these,
we added ESO archive GIRAFFE-FLAMES observations acquired with the setup HR15 
on May 27, 2003 \citep{pris07}.

The selection of the GES targets   and the fiber allocation procedure 
have been performed 
 following common guidelines described in \citet{rand18}
 with the aim of  maintaining the homogeneity within the GES dataset. 
 %Briefly,
%an unbiased large sample of candidate members with magnitudes down to  V=19\,mag, for observations with 
%the GIRAFFE spectrograph, and a smaller sample of brighter and most likely members,
% down to V=16.5\,mag, for observations with UVES, 
Targets were  
selected from the color-magnitude diagram (CMD)
 using optical BVI photometry from \citet{pris05}, 
VPHAS+ photometry from \citet{drew14}, and 2MASS astrometry.

Data reduction was performed at 
the Cambridge Astronomy Survey Unit (CASU) for GIRAFFE and 
at Arcetri for UVES \citep{sacc15} spectra, 
from which  radial and rotational velocities were derived. 
Spectrum analysis  and parameter homogeneization were described in \citet{lanz15}.
%includes as first step the determination of
%fundamental stellar parameters, i.e. effective temperatures and gravities, 
%and subs equently the determination of chemical abundances and other parameters, from
% the homogenization  of the results obtained by different GES nodes, perfomed by
%the  Working Group 15 (WH15) \citep{lanz15}.

The homogenized values constitute the official
 data set, the most recent internally distributed
through the   data release GESiDR5, as approved by the Working Group 15 (WG15).
 It includes the radial  velocities (RVs), rotational velocities v\,sini, effective
temperatures, gravities, $\gamma$ indices \citep{dami14}, lithium equivalent width (EW(Li))  and several
parameters of the H$\alpha$ line, as for example the Full Width at 
Zero Intensity (FWZI).
Effective temperatures and gravities derived by the Palermo 
 GES node were used for the
 objects not included in the WG15 recommended data set.
Using the 
sample of objects common to the two datasets,  we checked that they are in 
agreement. % per i numeri dettagliati vedere output param_dataset
As in the case 
of the Carina Nebula \citep{dami17b}, several spectra in our sample 
present unexpected features that are not considered in the WG15 
standard procedure, but that are instead dealt with in the procedure adopted
in the WG12 dedicated to PMS stars. In this case, RVs are
 provided by the Catania   \citep{fras15},
 and   Palermo GES nodes \citep{dami14}.
For this reason, in general, 
we adopted the official WG15 RV values, but in the cases of  spectra with 
 features peculiar to young stars (e.g. nebular lines), we adopted the RVs released by 
 WG12, Palermo and Catania nodes, according to the best agreement among the values.
 
The adopted dataset  in the field of NGC\,6530 includes  spectroscopic values 
of 2077 stars, obtained by one or more spectra 
observed with several  setups of GIRAFFE and UVES, as detailed in Table\,\ref{setuptab}.
The sample also includes   335 spectra from the ESO/FLAMES archive, reduced and analyzed 
homogeneously to the GES spectra. 
\begin{table}
\caption{GIRAFFE and UVES FLAMES setups used for the GES observations of NGC\,6530. The corresponding number
of stars observed with these setups is reported in Column\,2. Spectra acquired with HR15 
were taken from the ESO/FLAMES archive.}             % title of Table
\label{setuptab}      % is used to refer this table in the text
\centering                          % used for centering table
\begin{tabular}{c c}        % centered columns (2 columns)
\hline\hline                 % inserts double horizontal lines
Setup & \# stars \\    % table heading
\hline                        % inserts single horizontal line
%see ../MACRO/output of collect_data_ges.pro
HR14A--HR3--HR5A--HR6                                                                   &          11\\
HR14A--HR6                                                                            &           1\\
HR15                                                                                 &          85\\
HR15N                                                                                &        1924\\
HR15N--U520                                                                           &           7\\
U520                                                                                 &          33\\
U520--U580                                                                       &           7\\
U580                                                                                 &           9\\
\hline                                   %inserts single line
\end{tabular}
\end{table}
The spatial distribution of these objects is shown in the upper/left panel
of Fig.\,\ref{plotacis}.
  
%IDL> ind=where(strtrim(par_wg15.ges_fld,0) EQ 'NGC6530' and par_wg15.setup EQ 'U580 ',cind)           
%IDL> help,ind
%IND             LONG      = Array[16]
%IDL> ind=where(strtrim(par_wg15.ges_fld,0) EQ 'NGC6530' and par_wg15.setup EQ 'HR15N',cind)  
%IDL> help,ind                                                                               
%IND             LONG      = Array[1935]
%IDL> ind=where(strtrim(par_wg15.ges_fld,0) EQ 'NGC6530' and par_wg15.setup EQ 'HR15 ',cind) 
%IDL> help,ind                                                                               
%IND             LONG      = Array[335]

%\subsection{Optical  data\label{opticaldata}}

\subsection{Optical and NIR photometry}
Several deep optical and NIR astrometric/photometric catalogs in the region 
of NGC\,6530 are available in the 
literature:
\begin{itemize}
\item{BVI photometry (down to V$\sim 23$) obtained from 
the Wide Field Imager (WFI) at the MPG/ESO 2.2\,m Telescope  
within a field of view (FOV) of 34\min $\times$ 33\min\, \citep{pris05}. 
This catalog includes  a total of 53\,581 objects, taking also into account  the bright sources
found by \citet{sung00}.  
The spatial distribution of the WFI sources is shown 
in the upper/right panel of Fig.\,\ref{plotacis}.}

\item{ugriH$\alpha$ VPHAS+  \citep{drew14} photometry described in \citet{kala15}. In the
region covered by the GES data this catalog includes 94\,826 sources.
%help,where(vphas.ra GT 270.6 and vphas.ra LT 271.6 and vphas.dec GT -24.6 and vphas.dec LT 24.08)
%<Expression>    LONG      = Array[94826]
Their spatial distribution is shown in the lower/left panel of Fig.\,\ref{plotacis}.
 }
\item{JHK NIR photometry taken from the All-Sky Point Source public catalog of the Two Micron All Sky Survey (2MASS) \citep{cutr03}.}
\item{3.6, 4.5, 5.8, and 8.0\,$\mu$m bands {\it Spitzer} IRAC photometry within a FOV of 
42\farcm5$\times$30\farcm0, from which 64   Class\,I/0 and 168 Class\,II sources have been
identified by \citet{kuma10}.}
\item{NIR/MIR spectral energy distributions (SED) excess sources defined in the MYStIX Probable Complex Members (MPCMs)
\citep{feig13,broo13}. Objects with infrared excess were 
identified by modeling their NIR/MIR SEDs as
 circumstellar dust in a disk or infalling envelope as done in \citet{povi11}.}
\item{parallaxes and proper motions from the \emph{Gaia} DR2 catalog \citep{gaia16a,gaia18,lind18}.}
\end{itemize}

\subsection{X-ray data \label{xdata}}
The NGC\,6530 region has been covered at high spatial resolution
by two X-ray Chandra ACIS-I observations. The first
one, indicated here as ACIS-I/A, was a 60\,ks observation carried out on June 
18-19th 2001,
and was centered on 
RA=18$^{\rm h}$04$^{\rm m}$24\fs38, 
Dec=-24\deg21\min05\farcs8. 
The X-ray analysis was presented in \citet{dami04}. 
The second one,  indicated here as ACIS-I/B, centered on 
RA=18$^{\rm h}$03$^{\rm m}$45\fs10, 
Dec=-24\deg22\min05\farcs0, was a 172\,ks observation,
obtained from the combination of three ACIS-I observations of 127, 15 and 30\,ks respectively.
%carried out on July 25th, 2003  (P.I. M. Gagn\`e).
%http://cda.harvard.edu/chaser/dispatchOcatResults.do
The FOVs of the two X-ray observations are drawn in the lower/right panel of Fig.\,\ref{plotacis}.
The two ACIS-I FOVs are partially overlapping and part of the second one covers a region outside the WFI FOV. 

The two observations were combined and a new list of 1510 X-ray detections has been derived
using the procedure described in \citet{dami04}.
The spatial distribution of these objects is shown in the lower/right panel of 
Fig.\,\ref{plotacis}.

\section{Cross-correlations among catalogs}
In this section we first describe  how we performed the cross-correlation
between the new X-ray catalog and the WFI/2MASS optical/NIR catalog. 
From this catalog, only objects observed in the Gaia-ESO survey program have been considered
 for the present analysis.
 
The number of X-ray sources falling in the WFI FOV is 1415. 
 %/us1/loredana/wfidata/oc31/ANALISI_2008/TABLES/matched_src.txt
 % INWFIX          LONG      = Array[1415] ; /us1/loredana/wfidata/oc31/ANALISI_2008/MACROS/plot_acis.pro 
%OUTWFIX; /us1/loredana/wfidata/oc31/ANALISI_2008/MACROS/plot_acis.pro 
The number of optical WFI sources found within the two ACIS-I FOVs is  14\,229. % WFIACIS 
To cross-correlate the 1415 X-ray sources with the objects in the
optical catalog, we used the  procedure described in \citet{pris05}
and we considered a variable matching radius  that takes the X-ray position error,
 $\sigma_X$ into account. 
We first used a matching distance $d=1\sigma_X$ and found a systematic shift of 
(RA$_{WFI}$-RA$_X$)=-0\farcs14  
%  RESTORE,'../SAVEFILES/oc31_wfi_sung_x_ir_newx.save',/v corr_ra
and (Dec$_{WFI}$-Dec$_X$)=0\farcs20 between the optical and X-ray positions of the matched objects.
%  RESTORE,'../SAVEFILES/oc31_wfi_sung_x_ir_newx.save',/v corr_de
We corrected the X-ray coordinates for this shift and cross-correlated again the 
two lists of coordinates,
using a matching distance $d=4\sigma_X$, with a minimum  matching distance
of  1\farcs5.  
%For the multiple identifications, we used a matching radius 
%equal to  $d=1.5\sigma_X$. 

With these conditions we have a total of 1\,178 matches, % MWFIX           LONG      = Array[1178]
%IDL> restore,'../SAVEFILES/oc31_wfi_sung_x_ir_newx.save',/v
%IDL> data=OC31_WFI_SUNG_X_IR_NEWX
%IDL> help,where(data.multix EQ 1)
%<Expression>    LONG      = Array[1043] /1 =1043  singole
%IDL> help,where(data.multix EQ 2)
%<Expression>    LONG      = Array[104] /2 = 52 doppie 
%IDL> help,where(data.multix EQ 3)
%<Expression>    LONG      = Array[18] / 3 = 6 tripla 
%IDL> help,where(data.multix EQ 4)
%<Expression>    LONG      = Array[8] /4 = 2 quadrupla
%IDL> help,where(data.multix EQ 5)   
%<Expression>    LONG      = Array[5] /5 =1  quintupla 
%IDL> help,where(data.multix EQ 6)
%<Expression>    LONG      =           -1
of which 
 1\,043 are X-ray sources with one optical counterpart,
 52 are X-ray sources with two optical counterparts,
 six are X-ray sources with three optical counterparts,
 two are X-ray sources with four  optical counterparts and 
 one  X-ray source with five optical counterparts. 
Therefore, the total number of X-ray sources with WFI counterpart(s) is  1\,104 %WFIX 
while the number of X-ray sources without an optical counterpart is 311. % 1415-1104

We considered all X-ray detections, including the 95 X-ray sources falling outside the WFI FOV, 
and cross-correlated them with the 2MASS catalog. Since the X-ray coordinates can be shifted with respect
to the 2MASS coordinates, we first cross-correlated the entire list of X-ray sources with the 2MASS
All-Sky Point Source Catalog,  where we selected only objects obtained from aperture photometry or profile-fitting
(ph\_qual flag equal to 'AAA' or 'BBB' or 'CCC' or 'DDD'). Using the same procedure adopted for the match with the WFI
catalog, we first corrected for the systematic shift in the X-ray positions 
(RA$_{2MASS}$-RA$_X$)=0\farcs13  %  RESTORE,'../SAVEFILES/twomass_x.save',/v corr_ra,corr_de
and (Dec$_{2MASS}$-Dec$_X$)=0\farcs09. Using the new coordinates, we obtained a total of 462
2MASS counterparts % print,n_elements(twomass_x)
of 444 X-ray sources. %help,UNIQ(twomass_x.id_x,sort(twomass_x.id_x))
Among these, 53 X-ray sources  have 2MASS counterpart(s)  but no WFI counterpart.

%The 1178 WFI objects and the 53 2MASS sources with an X-ray counterpart are plotted in Fig....%\,\ref{radeacis}
%with different symbols.

We are left with a catalog of 54\,110 objects 
 %restore,'../SAVEFILES/oc31_wfi_sung_x_ir_newx.save',/v
including the 53\,581 WFI objects from \citet{pris05}, 123 optical 
sources from \citet{sung00} without WFI counterpart and  the 406 X-ray sources without optical counterpart
(only 53 of these last objects have a 2MASS counterpart). The 
GES targets with an X-ray counterpart are 471.
%restore,'../SAVEFILES/match_phot_ges.save',/v
%data=OC31_WFI_SUNG_GES 
%help,where(data.flag_x EQ 1) 
%<Expression>    LONG      = Array[471]

Finally, we cross-matched 
the  catalog of 2077 objects observed with GES with the  lists
of NGC\,6530 candidate members given by \citet{kuma10} and \citet{feig13}
and with the \emph{Gaia} DR2 catalog in this region. In conclusion, we found that
the list of stars  spectroscopically observed with GES includes:
i) 1\,733 objects with WFI-BVI photometry;
%restore,'../SAVEFILES/match_phot_ges.save',/v
%restore,'../SAVEFILES/collect_data_ges.save',/v
%wfi=OC31_WFI_SUNG_GES
% help,where(finite(wfi.v) and finite(wfi.i) and finite(wfi.b))
%<Expression>    LONG      = Array[1731]
ii) 1\,423 objects with VPHAS+ riH$\alpha$ magnitudes \citep{drew14,kala15}
%help,where(finite(udata_giraffe.hasmag) and finite(udata_giraffe.rsmag) and 
%finite(udata_giraffe.ismag))
%<Expression>    LONG      = Array[1423]
iii) 1976 objects with 2MASS counterparts;
%IDL> help,where(finite(udata_giraffe.j_2mass) and finite(udata_giraffe.h_2mass) and finite
%(udata_giraffe.k_2mass))
%<Expression>    LONG      = Array[1976]
iv) 50 (8) sources, classified as Class\,II (Class\,I/O)  YSOs, with {\it Spitzer} IRAC
magnitudes;
%IDL> help,where(finite(irac_ges.__3_6_))
%<Expression>    LONG      = Array[58]
%IDL> help,where(finite(irac_ges.__4_5_))
%<Expression>    LONG      = Array[58]
%IDL> help,where(finite(irac_ges.__5_8_))
%<Expression>    LONG      = Array[58]
%IDL> help,where(finite(irac_ges.__8_0_))
%<Expression>    LONG      = Array[58]
%IDL> help,where(finite(irac_ges.__8_0_))
%<Expression>    LONG      = Array[58]
%IDL> help,where(irac_ges.cl EQ 'I')
%<Expression>    LONG      = Array[8]
%IDL> help,where(irac_ges.cl EQ "II")
%<Expression>    LONG      = Array[50]
v) 139 (23) objects classified as Stage\,O/I or Stage,II/III  (ambiguous) in the MYstIX project;
%RESTORE,'../SAVEFILES/read_mystix.save',/v
%mystix=where( mystix_ges.st EQ 1 or mystix_ges.st EQ 2)                      
%IDL> help,mystix                                            
%MYSTIX          LONG      = Array[139]
%IDL> mystix=where( mystix_ges.st EQ -1)                     
%IDL> help,mystix                       
%MYSTIX          LONG      = Array[23]
vi) 2013 objects with a counterpart in the \emph{Gaia} DR2 catalog  with 
  proper motions (PM) and parallaxes.
% help,where(finite(gaia_ges.parallax))     
%<Expression>    LONG      = Array[2013]

\section{Membership strategy}
The simultaneous use of spectroscopic data with literature optical and NIR photometry and X-ray data
can be exploited to study the membership in very young clusters, at a very accurate level, allowing us 
to maximize the number of true  positives, i.e. the confirmed members, 
and minimizing the number of false positives, i.e. the contaminants. This is crucial
for selecting a sample as  complete as possible, since
the combination of the available indicators
allows us to reject  true negatives (non members) and to retain as many as possible false negatives,
that are  genuine members passing some criteria that are  negatives for other criteria. 

%the membership is  not positive for a given criterion 
%(for lack of information or if that criterion
% is not sensitive at some spectral types) but that can be members for other criteria.

For example, there are strong youth criteria such as high values of  
EW(Li) or  IR excesses 
and broad H$\alpha$ emission lines,
 signatures of circumstellar disks and accretion/outflow, respectively, that
are very useful to select confirmed members (true positives). In fact, it is
  very unlikely that they select contaminants (false positives) since the timescales typical of  disks and accretion are
 relatively low (smaller than few 10 Myr).
Nevertheless, there are objects that can be 
  negative for some criteria  and positive for others. For example, objects with 
 EW(Li)  smaller than the adopted threshold
 or without IR or H$\alpha$ line excesses, can be either
 early type or lithium depleted young stellar objects   \citep[YSOs--][]{pall07,sacc07}  or even 
 YSOs without circumstellar disk and/or accretion processes.   

Other criteria, such as the X-ray emission from young stars are 
known to decay on longer
timescales, depending on the spectral type \citep{jeff14,pris16}, and therefore,
 if all X-ray detected 
objects are considered young stars, they are likely to
include a low but not negligible fraction of false positives, since there is a 
chance to find $\sim$100-200 Myr old field stars or older close binary stars,
 with X-ray emission, that are unrelated to the young cluster \citep{wrig10}.
 Even for X-rays, false negatives can be found,
depending on the spectral types and/or the observational limits.

The RV membership criterion,  by selecting the stars with RV 
around the cluster mean RV,
 allows us to include a large fraction of true positives but,
  for this indicator, the chance to include also
false positives is not negligible. On the other hand, false negatives, that are genuine
members with RV outside the cluster RV range can be found, if the objects are binaries or if the RV is not
well determined as in case of artifact lines in the spectra with
 uncorrected sky subtraction due to the strong nebular contribution 
 \citep[see][]{dami17b}.
A further kinematic membership criterion  is provided by 
\emph{Gaia} DR2 PMs and parallaxes, since they allow us to select  stars with a common  motion 
and located at similar distances.

Finally, the $\gamma$ index defined by \citet{dami14}, being an indicator of the stellar gravity,
is a very useful criterion to select true negatives. 
In fact, \citet{dami14} showed
that  for late spectral type stars, the $\gamma$ index of 
giants is significantly larger than 
that of PMS stars, that is only slightly larger that that
 of MS stars. For this reason, the
 $\gamma$ index can be used to select giants and then,
to discard them as true negatives from the sample of YSOs.

In order to exploit all available criteria, 
and take into account for both their potential and limitations,
in  this work we adopted the following membership strategy. We first   defined as candidate members
all the objects that are positives  in at least two of the following membership criteria:
RV,  EW(Li), H$\alpha$ FWZI, photometric r-H$\alpha$ color excesses  \citep{kala15},
X-ray detections, NIR/MIR color excesses, {\it Spitzer} IRAC color
 excesses \citep{kuma10}
and \emph{Gaia} DR2 PMs and parallaxes. 
Finally, we discarded all expected true negatives
based on the $\gamma$ index. 

With this strategy we are confident of maximixing the number of certain members (true positives),
and discarding as many as possible contaminants (false positives),
and not to discard a priori genuine members that are negatives to one or more criteria (false negatives)
and reject  only non members (true negatives).
In fact, the probability of selecting false positives simultaneously  using more than one criterion 
(even only two) is definitively lower than that of using only one criterion. At the same time, 
using simultaneously two membership criteria, without a priori discarding (false or true) negatives
for all the criteria, %, with the only exception of the $\gamma$ index, 
allows us to not discard genuine members, leaving us the opportunity to study  the global  properties 
of the cluster. In addition, this strategy reduces the bias due to the lack of youth or membership
indication in one or more criteria, due to the observational limits.

In the following subsections, we detail  the selection
 of the candidate members that includes  in this step both true and false positives,
 for each of the adopted criteria. 

\subsection{Radial velocities}
\label{radvelsect}
The first step to select candidate cluster members by their RV is to study the RV distribution of a very reliable
sample of  cluster members (only true positives).
This is crucial to derive the shape of the cluster RV distribution and therefore its statistical properties.
The  sample suitable to model the cluster RV distribution
 has been chosen from a sample of filtered  members where  all possible negatives
were discarded as described in \citet{rand18}.
The maximum likelihood technique adopted to model the observed RV distribution 
has been recently described for other clusters in \citet{rand18} and it has been previously adopted 
for other clusters included in the GES project \citep{jeff14,sacc15}. The technique 
 takes into account  the binary contribution \citep{cott12} and the uncertainty distribution 
\citep{jack15}. 
In addition, to take into account  the fraction of contaminants included in the starting samples,
the model consists of two Gaussian components, one for the cluster and one broad Gaussian for the
background of the contaminating stars.
In the case of NGC\,6530, the best fitting parameters we found are 
% see email by Jackson 19 giugno 2017 Re: NGC6530 analysis with DR5 data
RV$_{\rm cl}$=0.17 km/s and $\sigma_{\rm cl}$=2.42 km/s for the cluster and 
RV$_{\rm fld}$=-10.69 km/s and $\sigma_{\rm fld}$=32.17 km/s for the field stars.
The fraction of stars belonging to the cluster population is f$_{\rm cl}$=0.44.
A more detailed study of the kinematics and RV distribution of Lagoon Nebula is presented 
in Wright et al. (2018, submitted).

Based on these results, we consider the  RV$_{\rm cl}$ and $\sigma_{\rm cl}$ as representative for the
cluster RV distribution and we consider as candidate members, positives with respect to RV criterion,
all the objects with RV around   RV$_{\rm cl}$ and within 5$\sigma_{\rm cl}$, 
corresponding to 
a probability lower than 0.57 parts per million\footnote{assuming the cluster member RVs
are distributed in a Gaussian fashion} 
of finding cluster members outside this range.
The true probability is  larger than this value
since the uncertainty distribution has extended tails
that are better represented by a Student-s t-distribution than a by normal distribution \citep{jack15}.
With this criterion, we selected \memrv RV candidate members. % see membership_rv.pro
%A detailed kinematic study of RVs in the Lagoon Nebula is the subject of an
%ongoing paper by Wright et al. (2018, in preparation).
%restore,'../SAVEFILES/membership_final.save'
%help,where(membership.rv_mem EQ 1)    
%<Expression>    LONG      = Array[822]
\subsection{Lithium equivalent width}
Theoretical models predict that stars with ages younger
 than 10\,Myr have cosmic abundances of  lithium
\citep{bara98,sies00}.
% As already shown in \citet{pris07}, the EW(Li) in 
%NGC\,6530 young stars  is temperature
%dependent. 
In order to establish a EW(Li) threshold suitable to include a list as complete as possible of members, we considered the EW(Li) distribution as a function of the effective temperatures  of the 
RV candidate members, mostly formed by cluster members. 
We found that most of the potential members
have EW(Li) $>200$\,m$\AA$ for stars with T$_{\rm eff}\lesssim 5200$ K.
For  stars with T$_{\rm eff} \gtrsim 5200$ K, the EW(Li) of possible members is 
 in the range [100-200]\,m$\AA$, even though, at these temperatures, the lithium strength 
is  no longer a sensitive age indicator. 
As mentioned  before, we cannot discard a priori the presence of lithium depleted (or partially depleted)
 members with T$_{\rm eff}\lesssim 5200$ K, as found in other young star forming regions 
\citep{pall07,sacc07}. Therefore, we adopted a  conservative 
threshold of 100\,m$\AA$, independently of the effective temperature, and 
defined as EW(Li) candidate members all  objects with EW(Li) larger than this value. Using this relatively low
threshold, we are aware of the risk of  including  also false positives (many of which 
 will be discarded  with  the final
selection) but, on the other hand, 
we  include most of the candidate members with T$_{\rm eff}\gtrsim 5200$ K
and potential depleted objects.
The observed stars with EW(Li)$>$100\,m$\AA$ are 545.
%help,where(membership.li_mem EQ 1)
%<Expression>    LONG      = Array[545]
\begin{figure*}
 \centering
 \includegraphics[width=\textwidth]{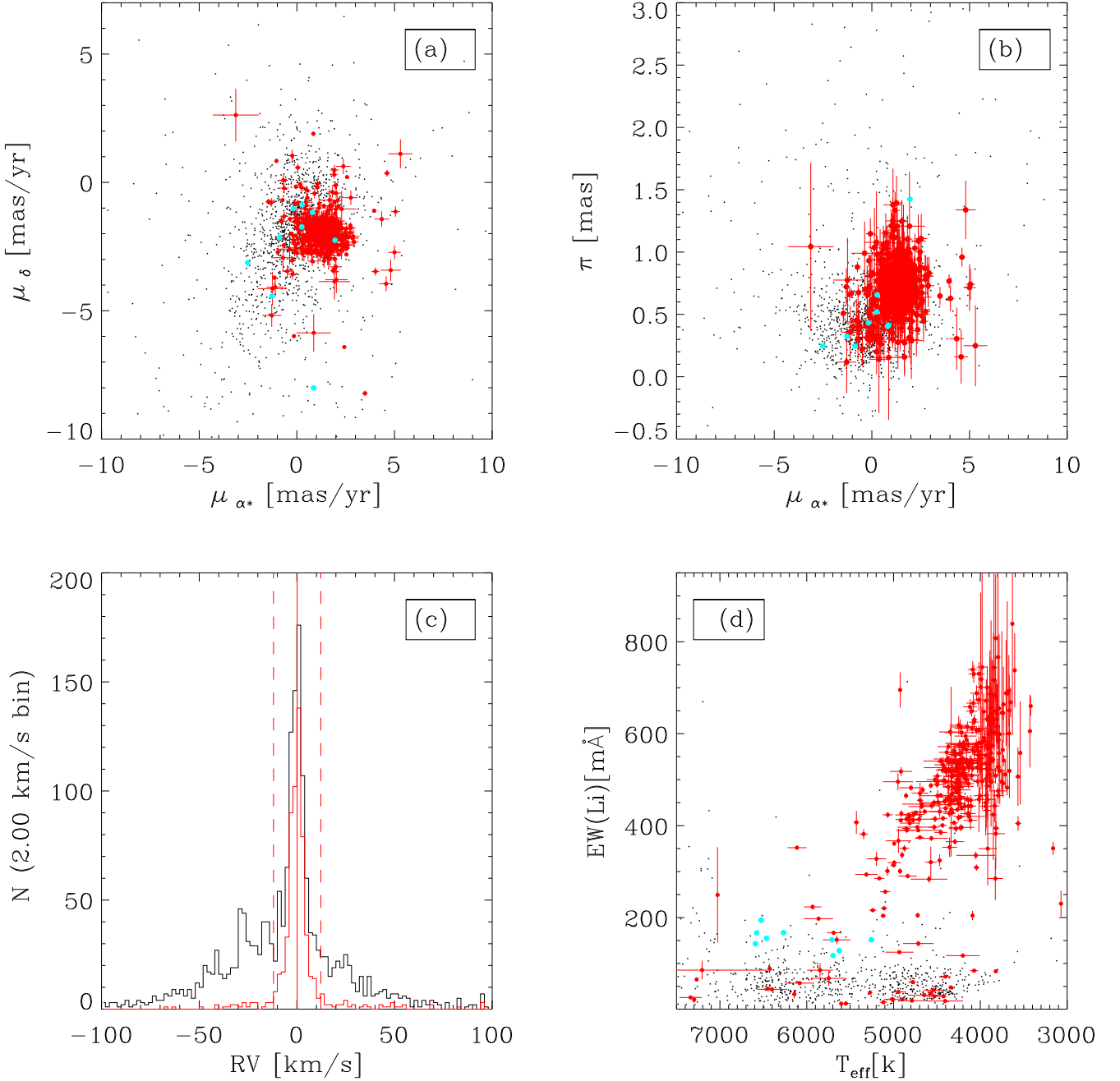} 
 \caption{\emph{Gaia} DR2 proper motion scatter plot (panel a), \emph{Gaia} DR2 
 parallaxes vs.
 $\mu_\delta$ (panel b), RV distribution (panel c), 
and EW(Li) as a function  of  the effective temperature (panel d) 
of all GES targets (black symbols).
 Objects selected as confirmed (probable) cluster members are indicated as red 
 ( cyan) dots.
The red histogram in panel c is the RV distribution of the confirmed cluster members, while dashed
red lines are the limits used for the selection of RV candidate members.}
\label{membershipplots}
 \end{figure*}

\subsection{H$\alpha$ line\label{halphasect}}
The H$\alpha$ line in spectra of young stars is a strong signature of chromospheric activity,
producing emission in the core of the line, or of circumstellar accretion and outflow,
 producing
emission of broadned lines due to high velocities of the gas when it impacts the stellar
surface \citep{boni13,pris16}.
Chromospheric activity  can be quantified through  the  H$\alpha$ equivalent width 
(EWHaChr) values from the GES recommended parameters  or the $\alpha_c$ indices
(measuring flux in the line core) defined in \citet{dami14}, while 
accretors can be selected 
 from the H$\alpha$ 10\% width or from the FWZI,  used, for example, in \citet{pris07} and \citet{boni13}. 
In the case of a young cluster such as NGC\,6530, surrounded by an \hii region, the 
H$\alpha$ line in the  observed stellar spectra  includes
also a significant contribution from the surrounding nebulosity.
 Due to the complex dynamics of the 
 ionized nebular gas \citep{dami17a}, 
  this contribution cannot be rigorously  quantified and subtracted
 with the traditional technique of the mean sky, 
estimated by taking a number of sky spectra from the same
region. In fact,
  sky subtraction can be over or underdone (Bonito et al., in preparation),
 and therefore, all measurements involving the peak of the H$\alpha$ line, 
 such as the H$\alpha$ 10\% or equivalent  widths, the EWHaChr and the $\alpha_c$ 
 indices are not reliable to quantify the stellar H$\alpha$
contribution. Nevertheless, while the accretion process implies a significant  line broadening,
the nebular emission typically affects only the line core. Therefore, spectra of accretors can be distinguished from
those of non-accretors with chromospheric activity and/or sky nebular emission,
 using the  H$\alpha$ line FWZI, that is the only measurement independent of line peak intensity.  
We used the GES recommended parameter FWZI and selected as candidate accretors all the objects that,  within 
 errors,\footnote{i.e. they are only compatible to belong to the population of accretors} 
have FWZI $>$ 4$\AA$.
This threshold   has been chosen because the objects with NIR (from JHK or  
{\it Spitzer} IRAC magnitudes) or  VPHAS+ r-H$\alpha$ excesses have FWZI $\gtrsim$ 5$\AA$.
With this condition, we selected 241 candidate accretors.
%help,where( membership.fwzi_mem EQ 1)                            
%<Expression>    LONG      = Array[241]
To this sample we added  31 of the 235 accretors defined in \citet{kala15}  using the r-H$\alpha$ excesses, that are  included in the GES sample.
%help,where( membership.vphas_mem EQ 1)
%<Expression>    LONG      = Array[31]

\subsection{X-ray and NIR/MIR membership}
%Candidate members from X-ray detections and NIR excesses
Among the 2077 stars observed within GES, there are 471 objects also detected in the X-rays and
%help,where( membership.x_mem EQ 1)                               
%<Expression>    LONG      = Array[471] 
243 objects with IR excesses in the JHK colors, according to the extinction-free indices defined in 
%help,where( membership.irex_mem EQ 1)
%<Expression>    LONG      = Array[243]
\citet{pris07}. In addition, we considered also the 50 Class\,II and the 8 Class\,0/I stars
defined in \citet{kuma10} 
%help,where( membership.irac_mem EQ 1)
%<Expression>    LONG      = Array[58]
and the 162 YSOs selected in the MPCMs project \citep{feig13,broo13},
%help,where( membership.mystix_mem EQ 1)
%<Expression>    LONG      = Array[162]
that are included  in the GES sample. We considered these  samples
as candidate cluster members.

\subsection{Kinematic membership with \emph{Gaia} DR2}
To define the range of PMs and parallaxes of candidate cluster members,
we first considered a fiducial sample of members confirmed by at least three of the
membership criteria described in the previous subsections.
For this subsample, we computed the median of PMs 
$\mu_{\alpha^\ast}$\footnote{we use the notation $\mu_{\alpha^\ast}=\mu_\alpha cos \delta$}
 and $\mu_\delta$ and parallaxes $\pi$ that are
1.34\,mas/yr, -2.03\,mas/yr and  0.74\,mas, 
respectively and the relative standard deviations that are
1.34\,mas/yr, 0.93\,mas/yr and 0.45\,mas, respectively. 
%These values are in agreement with those found in \citet{kuhn18}.
% see membership_gaia

We considered candidate members those within the two ellipses  in the planes
$\mu_{\alpha^\ast}$ vs. $\mu_\delta$ and $\mu_{\alpha^\ast}$ vs. $\pi$, where the
semi-minor and semi-major axes  were defined by the 1$\sigma$ values  
 of the fiducial sample of cluster members.
With these criteria we selected 516 candidate members from 
\emph{Gaia} DR2  data among the 2077 
GES targets.
\emph{Gaia} DR2 data together with GES RVs are analysed
 in Wright et al. (2018, submitted). 

\subsection{Final membership for GES targets\label{finalmem}}

By combining the previous criteria, we selected  \memall objects 
%restore,'../SAVEFILES/membership_final.save',/v
%help,where( membership.fin_mem EQ 1)   
%<Expression>    LONG      = Array[652]
%IDL> help,where( membership.fin_mem EQ 2)
%<Expression>    LONG      = Array[9]
that are included in at least two of the 
candidate member samples previously defined and have $\gamma$ index smaller than 1.018.
This latter condition has been applied to reject non-members, with gravity (inferred from the
$\gamma$ index)
inconsistent with PMS stars. We consider {\it probable members} the \memprob objects
with T$_{\rm eff}>5200$\,K for which 
the EW(Li) is in the range [100, 200]\,m$\AA$  since in these cases the EW(Li) is not a strong age indicator
and therefore for these objects the membership is positive only for one criterion. 
The remaining \memconf objects
are classified as {\it confirmed members}.

 Among the \memall confirmed or probable members, \memctts have been classified 
as classical T Tauri stars with {\it excess} (CTTSe), 
%help,where(( membership.disk EQ 1 or membership.accretor EQ 1) and membership.fin_mem GE 1) 
%<Expression>    LONG      = Array[337]
including accretors, on the basis of the H$\alpha$ FWZI or the r-H$\alpha$ colors,
 and/or objects with circumstellar disk, on the basis of 
 NIR excesses. The remaining \memwtts
%help,where(( membership.disk NE 1 and membership.accretor NE 1) and membership.fin_mem GE 1)
members have been classified as weak T Tauri stars with {\it photospheric} emission (WTTSp) 
including members without evidence of accretion and  
 without evidence of circumstellar disk.
 Literature and Gaia-ESO survey parameters of the selected cluster members
 are given in Tables\,\ref{tablemem1} and \ref{tablemem2}. 
Figure\,\ref{membershipplots} shows the $\mu_\delta$ vs. $\mu_{\alpha^\ast}$ and
the  $\pi$ vs. $\mu_{\alpha^\ast}$ scatter plots of all objects observed with GES.
Stars selected as confirmed or probable members are highlighted with different colors.
The RV distribution and   
EW(Li)  as a function of the effective temperature
 of the \memall objects selected as confirmed or probable members are also shown.
 There are few objects with proper motions and/or parallax values outside from the 
 typical cluster values. We checked that most of them are objects 
 with excess noise $>$1.0\,mas % limit used by Kuhn18
 and/or $\chi^2>800$  \citep{lind12} and/or with parallax errors $>$0.3, % see membership_plots.pro 
 that means they could be binaries or objects for which the 
 \emph{Gaia} DR2 kinematic parameters
 are not well determined due to confusion caused by the large stellar density towards NGC\,6530.

%\subsection{Additional candidate members not observed with GES}
%TBD\\
%X-ray detections not observed with GES
%dearrossare questi membri usando E(B-V) medio dei target GES vicini
%magari dimostrando che E(V-I) e' costante nelle diverse zone spaziali.

%Fare il diagramma HR  delle stelle  X con controparte ottica e vedere quelle con IR in modo da
%selezionare CTTS e non. (stabilire la contaminazione delle X-only WTTS dai membri selezionati con RV)

\section{Candidate giants and MS field stars in the GES sample\label{giantsandms}}
\begin{figure*}
 \centering
 \includegraphics[width=\textwidth]{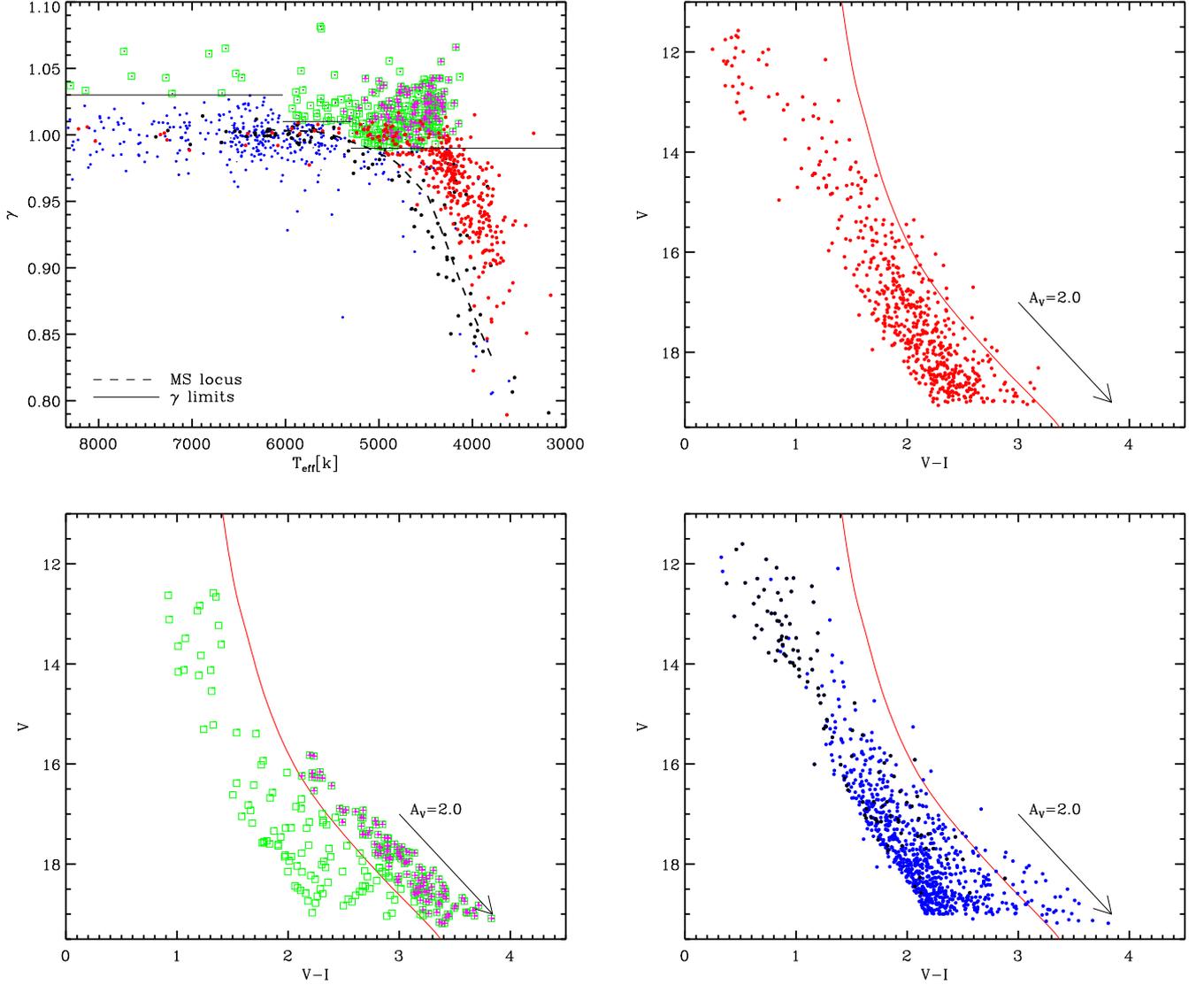} 
	\caption{ $\gamma$ index as a function of  effective temperature 
(upper/left panel) and observed V vs. V-I diagrams
of the  GES targets selected as cluster members 
	(red bullets),
	candidate giants (green squares), RC giants (magenta plus symbols),
	foreground (black bullets)  and background (blue bullets) MS stars. 
The horizontal solid segments  and the dashed line in the upper/left panel
indicate the limits used to select giants and the MS locus, respectively. 
The red solid line in the other panels is the 0.1\,Myr PISA isochrone 
	described in the text.}
\label{giantsplots}
 \end{figure*}

GES effective temperatures of field stars, in combination with the observed colors, can be used to 
derive the interstellar reddening of these objects and therefore to study the properties
  of the dust component of the Lagoon Nebula.

The sample of non-members includes both MS stars and giants,
these latter expected to be found mostly at distances significantly larger than 
the cluster distance,
i.e. beyond the Lagoon Nebula, while MS stars are expected to be found 
 both in front of and behind
 the nebula.

We describe in this section the use of the $\gamma$ index 
defined by \citet{dami14} to define candidate giants
within the GES sample of cluster non-members.
Figure\,\ref{giantsplots} shows the $\gamma$ index as a function 
of the effective temperature
for all the objects classified as members and non-members. 
The dashed curve indicates the locus of
MS stars from \citet{dami14}. Giant stars 
have lower gravities and therefore $\gamma$ indices higher than this limit.
In addition, red giants are expected to have
effective temperatures T$_{\rm eff}<6030$\,K, 
based on the \citet{bess88} relations.  Very hot giants are quite uncommon. 
Based on these considerations, we selected
as candidate giants all the objects with $\gamma$ indices higher than the horizontal limits shown in
 Fig.\,\ref{giantsplots}. The limits are temperature dependent to ensure the selection of objects
with $\gamma$ index larger than the MS locus.
In particular, we adopted a lower threshold ($\gamma >0.99$)
  for T$<$5300\,K, where the $\gamma$ index is more sensitive to the gravity 
  and  the separation between   MS stars and giants is larger while, for 
 5300$<$T/K$<$6030,  we used the limit $\gamma >1.01$.  
We note that  for 4700$<$T/[K]$<$5300 and $\gamma >0.99$,
many candidate giants    have $\gamma$ values similar to those of
 objects selected as  young cluster members.
 This occurs since, as predicted by theoretical models,
 in this range of effective temperatures,
 the ranges of gravities of
 giants and PMS stars younger than few Myr overlap
   \citep[see Fig.\,4 of ][]{dami14}.
 We included in the giant sample also the
few objects with T$>$6030\,K and $\gamma >$1.03.
%This sample includes 235+33+11=279  % gig1 and gig2 and gig4 output select_giants
This sample includes \giga+\gigb+\gigc=\gigtot
candidate giants that are shown as green squares in the Fig.\,\ref{giantsplots}.

 The CMDs of cluster members and candidate giants are shown in
Fig.\,\ref{giantsplots} (upper/right and lower/left panels, respectively).
Most of the selected giants fall  outside of the expected PMS region
in the V vs. V-I diagram,  distributed along the reddening vector
direction. As already noted in \citet{pris05}, these objects are dominated by
 very distant and reddened
 Red Clump (RC) giants \citep{gira99}.
 Among the sample of  \gigtot  giants, we selected the subsample of RC giants,
as the objects with V-I colors redder than the 0.1\,Myr isochrone computed 
 using the updated version of the PISA stellar evolution code 
\citep[see e.g.][]{rand18,togn18}.
The isochrone has been positioned 
at the cluster distance of 1250\,pc \citep{pris05}, 
assuming a mean reddening E(B-V)=0.3 and
the reddening law of R$_{\rm V}$=A$_{\rm V}$/E(B-V)=5.0. With this condition, we find that 
\gigrc % samp4 of select_giants
of the selected objects are RC giants.  

While we are  confident that these \gigrc are confirmed giants, the remaining 
sample of \gignorc %(281-117)
 candidate giants could be contaminated by MS stars, since
for early type stars,  the $\gamma$ index is not very sensitive in separating
objects with different luminosity classes. 
We consider all the remaining GES objects as the field MS stars,
i.e. those that are neither classified as cluster members nor giants.
 The CMD of MS stars is shown in Fig.\,\ref{giantsplots} 
(lower/right panel).
%RESTORE,'../SAVEFILES/intrinsic_colors_pisa.save',/v
%help,where(membership.fin_mem EQ 613
%<Expression>    LONG      = Array[161]
% help,where(membership.fin_mem EQ 615)
%<Expression>    LONG      = Array[117]
%help,where(membership.fin_mem GE 61) 
%<Expression>    LONG      = Array[280]

\section{Interstellar reddening\label{intred}}
%Using NGC2264, Gamma Vel, NGC2547 and NGC2516 ad calibrators to test if the
%col.-temperature relation depend on gravity and then on age. 
%There is an interesting discussion on this topic in \citet{sung13}.
%There are several references on the adopted transformations in the literature in Sect. 4.5
To convert GES effective temperatures
and surface gravity data to colors, we computed transformations
using the synthetic spectra library by
 \citet{cast03} 
for T$_{eff}>$6500\,K and the  \citet{alla11} ones for T$_{eff}<$6500\,K, integrated over 
 the adopted filter response. For  giants, we used the \citet{bess88}
intrinsic color transformations. 

To derive the interstellar reddening, we   used
 the  V-I colors, since the V and I magnitudes
are those where additional non-photospheric contributions, 
due to dust in the circumstellar disk, 
or blue excesses, due to  accretion, are negligible.

We only used  optical WFI photometry rather than the VPHAS+ photometry since the 
members (probable and confirmed) for which both  effective temperatures and the WFI V-I  (VPHAS+ r-i) colors
are available are in total \teffwfi (\teffvphas), with \teffwfivphas common objects within the two samples.
In addition, the anomalous reddening law adopted in this work 
for the cluster is defined in the Johnson-Cousin system.
%restore,'../SAVEFILES/collect_data_ges.save'
%RESTORE,'../SAVEFILES/vphas.save',/v 
%wfi=OC31_WFI_SUNG_GES_KAL
%aa=where(membership.fin_mem GE 1 and membership.fin_mem LE 2 and $
%	  finite(upar_lore.teff) and finite(udata_giraffe.rsmag-udata_giraffe.ismag))
%bb=where(membership.fin_mem GE 1 and membership.fin_mem LE 2  and $
%	 finite(upar_lore.teff) and finite(wfi.v-wfi.i))
%aa= 364
% bb=395
%cc=where(membership.fin_mem GE 1 and membership.fin_mem LE 2 and $%
%	  finite(upar_lore.teff) and finite(udata_giraffe.rsmag-udata_giraffe.ismag) and $
%	finite(wfi.v-wfi.i))
%help,aa,bb,cc 

\begin{figure}
 \centering
 \includegraphics[width=8cm]{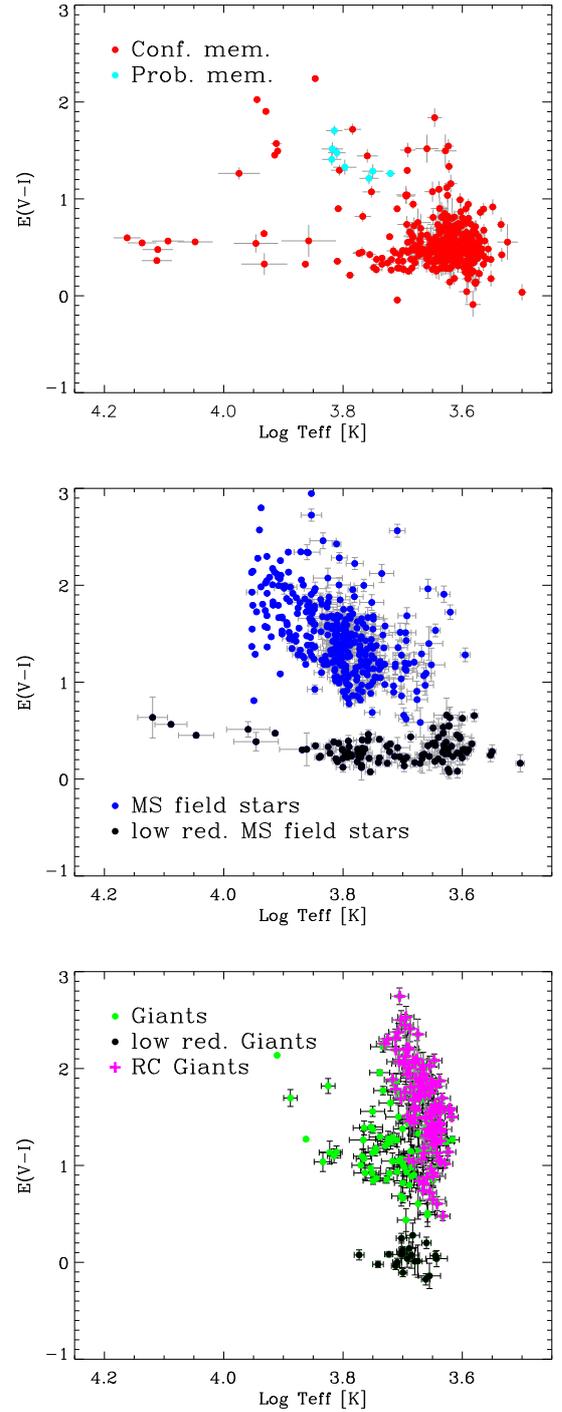} 
 \caption{Interstellar reddening E(V-I) vs. effective temperature 
 for cluster members, MS and giant field
stars.}
\label{intrinsiccolorspisa}
 \end{figure}

Figure\,\ref{intrinsiccolorspisa} shows  E(V-I) as a function of the 
effective temperatures for the
samples selected before, i.e. for cluster members, MS field stars and giants.
As expected, most of the cluster members (upper panel) share a similar reddening. 
\begin{figure}[!h]
 \centering
 \includegraphics[width=8cm]{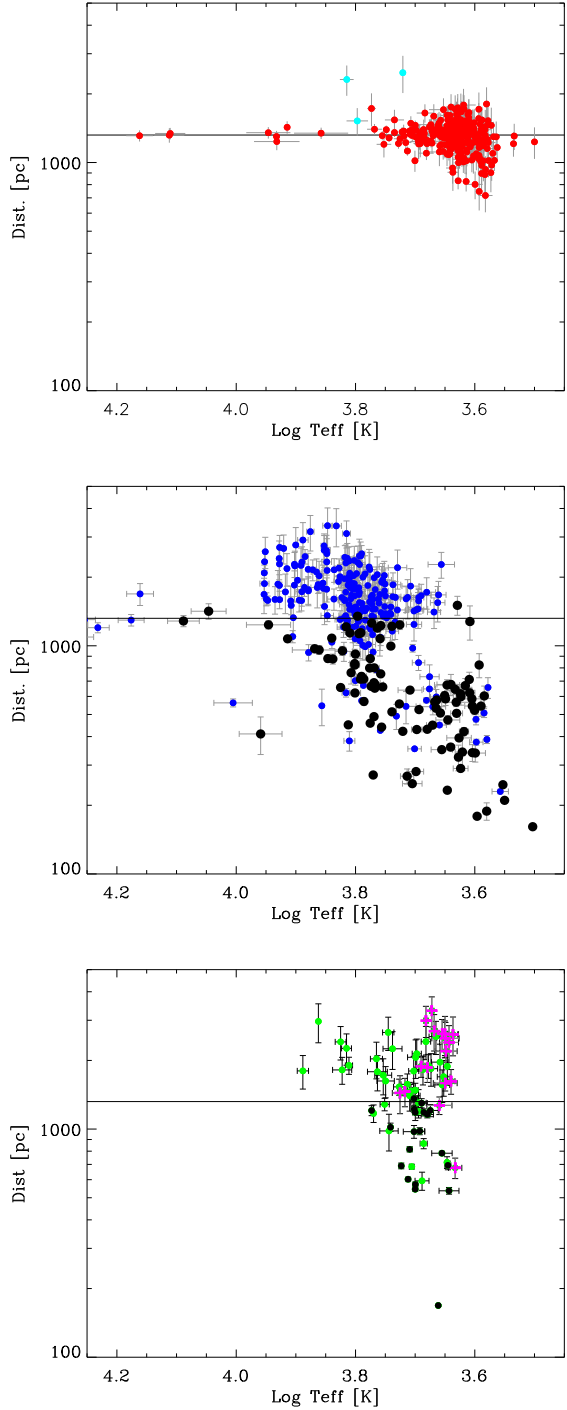} 
 \caption{Distances vs. effective temperatures for cluster members, MS and giant field
stars, derived using \emph{GAIA} DR2 data. 
Symbol colors are as in Fig.\,\ref{intrinsiccolorspisa}. Horizontal solid line indicates the cluster
distance.}
\label{parallteff}
 \end{figure}
% see intrinsic_colors_pisa output
%SAMPLE          LONG      = Array[652]
%LOWREDMEM       LONG      = Array[360]
%LOWREDMS        LONG      = Array[113]
%HIREDMEM        LONG      = Array[37]
%HIREDPBMEM      LONG      = Array[9]
%HIREDMEM_TWO    LONG      = Array[18]
%HIREDMEM_CONF   LONG      = Array[19]

% the following statistic reddening values are in the output of the intrinsic_colors_pisa.pro
In particular, there are 361 confirmed cluster members % lowredmem
 with E(V-I)$<$1. 
%restore,'../SAVEFILES/intrinsic_colors_pisa.save'
%aa=where(membership.fin_mem EQ 1 and data.evi LT 1.)
%<Expression>    LONG      = Array[361]
%print,mean(data[aa].evi,/NaN)
%IDL> print,mean(data[aa].evi,/NaN)
%     0.498852
%IDL> print,stddev(data[aa].evi,/NaN)
%     0.170646
%IDL> 
%IDL> print,mean(data[aa].ebv,/NaN)  
%     0.275268
%IDL> print,stddev(data[aa].ebv,/NaN)
%     0.168611
For this sample, the mean of the E(V-I) values is $<$E(V-I)$>$=0.50
and the standard deviation is 0.17, and  $<$E(B-V)$>$=0.27
and the standard deviation is 0.17. 
The $<$E(B-V)$>$ we found is in good agreement with that 
 given in the literature 
 \citep[E(B-V)=0.30--0.37, see e.g.][]{van-97,sung00,toth08}.  

  There are 26  members with E(V-I)$>$1. 
%help,where(membership.fin_mem EQ 1 and data.evi GE 1.)
%<Expression>    LONG      = Array[26]
Cluster membership is confirmed by more than two methods for 17 of them and by two methods
%help,where(membership.fin_mem EQ 1 and data.evi GE 1. and membership.nyes GT 2)
%<Expression>    LONG      = Array[17]
(RV and X-rays or RV and CTTS signatures or, in one case,  strong EW(Li) and CTTS signature) for 9 of them.
%help,where(membership.fin_mem EQ 1 and data.evi GE 1. and membership.nyes EQ 2)
%<Expression>    LONG      = Array[9]
 We will consider these objects as very reddened confirmed cluster members, since there is no reason
to discard them as members only because of their higher reddening. Their properties are discussed in the next sections.
In addition, there are further eight objects, classified as probable cluster members,
 with E(V-I)$>1$.
%help,where(membership.fin_mem EQ 2 and data.evi GE 1.)              
%<Expression>    LONG      = Array[8]

The E(V-I) distribution  of the sample of MS field stars is also shown
 Fig.\,\ref{intrinsiccolorspisa} (middle panel).
Even in this case, there is a sample of very low reddening MS stars 
(see also Fig.\,\ref{giantsplots}).
%see output intrinsic_colors_pisa
 Those with   E(V-I)$<0.5$
have a mean $<$E(V-I)$>$=0.31 ($<$E(B-V)$>$=0.22)
and  a standard deviation of 0.13 (0.14).
 These values, distributed within a very narrow reddening range,
 are slightly smaller than those found for the cluster,
and this suggests,  as expected,
 they are MS stars located in front of the Lagoon Nebula.
In contrast, in the sample of highly reddened MS stars,  
reddening values are spread over a larger range,
 0.8$\lesssim$E(V-I)$\lesssim$2.5, with a clear gap in the E(V-I) distribution
 corresponding to members. This suggests 
they are located at a larger distance,  and therefore, beyond the Lagoon Nebula.

We find a similar result for the sample of objects classified in this work as giants 
(bottom panel in Fig.\,\ref{intrinsiccolorspisa}).
There are few low reddening candidate giants with E(V-I)$<$0.2, expected to be nearby objects,
while most other giants suffer from large reddening. This latter sample
 includes all the RC giants selected  above,
with a trend with  the E(V-I) values, with hotter RC giants being more reddened than 
 cooler ones.
Again, we explain this trend as a distance effect,  the hotter giants being
 also  brighter,
and therefore visible at larger distance. A quite flat distribution of  E(V-I)  is, instead, found
for the non-RC giants with E(V-I)$\sim$1.1, likely located at a mean  distance smaller
than the RC giant distances,  not far beyond the Lagoon Nebula.

The correlation between the distance of the different samples with their reddening 
is confirmed by the \emph{Gaia} DR2 data. Figure\,\ref{parallteff} 
shows the distance derived from the \emph{Gaia} DR2
 parallax for the same samples shown in Fig.\,\ref{intrinsiccolorspisa} as 
 function of the effective temperatures.
For these plots, we selected only objects with relative errors in parallax $<20$\%.

Cluster members are distributed around a mean distance of 
\clustdist$^{\footnotesize +74.2}_{\footnotesize -66.7}$\,pc % see output cluster_distance.pro 
computed from the weighted mean parallax
equal to 0.7573$\pm$0.0403\,mas. 
This value has been obtained  using the fiducial sample including all the
objects that are members for at least three criteria.
The error on the parallax has been computed as
the error on the mean. To this, we added
 the systematic error equal to 0.04\,mas \citep{lind16}
estimated for \emph{Gaia} DR2 parallaxes.
 The value of the cluster distance we found is only slightly higher, 
 but in agreement within the errors, with the value of 1250\,pc derived 
 in \citet{pris05}. 
 %The value is also in agreement with 
 %the cluster distance of 1325\,pc found by Damiani et al. (2018, submitted),
 % by using \emph{Gaia} DR2 data in a larger region of the cluster.

 MS and giants affected by low reddening (E(V-I)$\lesssim$0.5)
are in front of the cluster, i.e. foreground field stars. 
The foreground MS stars show a quite linear
trend with the distances that increase towards  hotter stars 
(i.e. the most luminous stars).

In contrast, MS and giants affected by high reddening (E(V-I)$\gtrsim$1) 
are found behind the Nebula
and the cluster. This confirms that the reddening is strongly correlated 
with the distance 
and that the reddening gap both for MS stars and giants is due to the dust
 in the  Lagoon Nebula.

%We find that among the low reddened cluster members,
% hot members (T$_{eff}\gtrsim$5200\,k) are affected 
%by a reddening slightly smaller than cool members. We exclude that this trend is due to calibration
%issues in the GES effective temperatures at some spectral types. In fact, this trend is not 
%observed in the low reddened MS stars distributed at all spectral ranges covered by  the GES survey.
%A possible explanation of this results will be discussed in the following section. 

\section{Reddening law \label{redlawsect}} 
In a paper dedicated to the study of the interstellar extinction and its variations with wavelength,
\citet{card89} computed for the star Herschel\,36 an extinction law $R_V=A_V/E(B-V)=$5.3, i.e.
a peculiar value, higher than the standard one $R_V=$3.1, usually adopted for the Galaxy.
Since the work of \citet{walk57}, who first suggested an abnormal extinction law,
several authors have confirmed  this result, but at the same time, 
the analysis of other observations has favoured  a standard reddening law
 in the cluster NGC\,6530 \citep[see][for a review]{toth08}.

The dataset used in this work, including both  photometric and spectroscopic information,
can be exploited in several ways to investigate the 
reddening law  in this region. 

As a first step, we considered the sample of confirmed cluster members, classified 
in Sect.\,\ref{finalmem},
as WTTSp stars. This choice was made to ensure that the observed colors only depend  on
 interstellar reddening (and T$_{\rm eff}$).
In fact, peculiar objects such as the CTTSe   might
have colors that include other contributions due to  dust and/or 
 gas in their circumstellar disk. 

 \begin{figure}
 \centering
 \includegraphics[width=9cm]{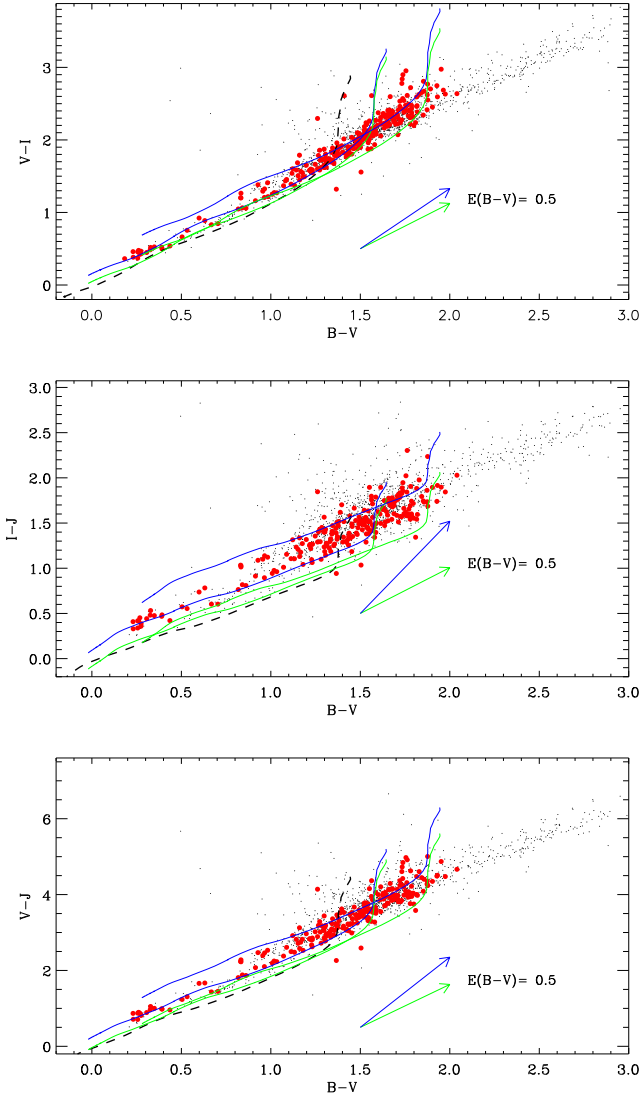} 
 \caption{Color-color diagrams of the GES targets (black dots) and WTTSp
  cluster members (red points)
for three different sets of color combinations. The 
dashed black line is the theoretical color-color locus
  adopted in this work (see Section \ref{intred}). 
  Green and blue solid lines are the  same locus
reddened by E(B-V)=0.2 and 0.5 assuming  $R_V=$3.1 and $R_V=$5.0, respectively.}
\label{membersredlaw}
 \end{figure}
 
  \begin{figure*}
 \centering
 \includegraphics[width=\textwidth]{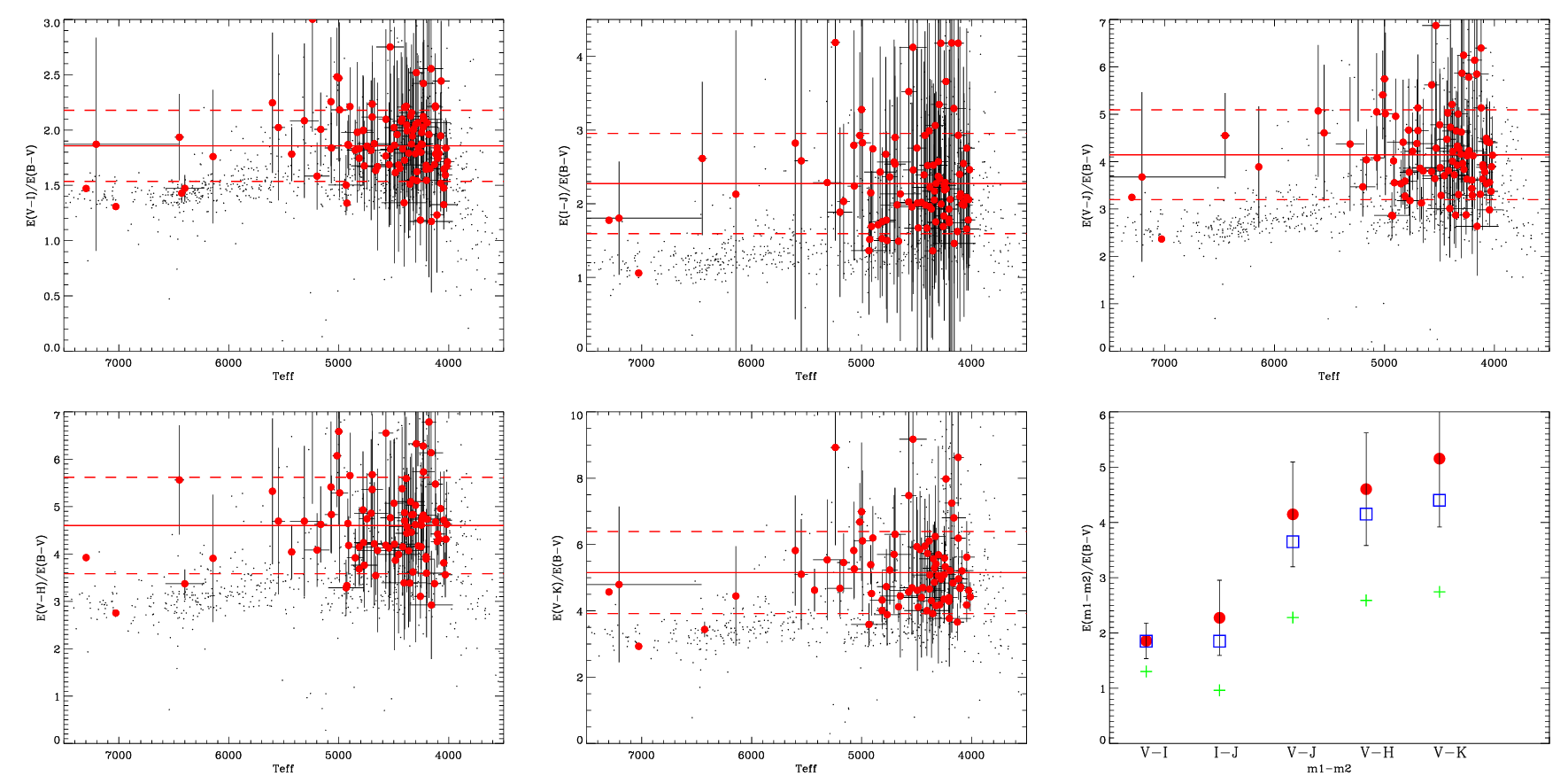} 
 \caption{First five panels: ratios of individual reddenings in several combinations of 
colors as a function of the effective
temperatures for all GES targets (small dots) and for WTTSp cluster members with 
T$_{\rm eff}>$4000\,K (red points). Bottom/right panel: mean ratios of reddening values for the 
WTTSp cluster members 
in several colors (red circles),  compared with the 
expected analogous values given for $R_V=$3.1 (green plus symbols) and $R_V=$5.0 (blue squares) 
based on the \citet{fior03} relations.}
\label{wttsratred}
 \end{figure*}

For the WTTSp cluster members, we compared several observed color-color diagrams with the expected unreddened 
intrinsic colors, as shown in Fig.\,\ref{membersredlaw}. By  shifting this locus along the
reddening vector for a suitable reddening law, a reasonable match between
the intrinsic reddened locus and the observations should be found.

Since most  WTTSp  in our sample have E(B-V) values between 0.2 and 0.5,
we shifted the adopted intrinsic relation assuming  these two values, with the aim of 
matching the bulk of the observed colors. 
As it is known, in the color-color diagrams shown in Fig.\,\ref{membersredlaw}, the intrinsic color-color locus is degenerate with respect to the 
reddening direction, except in the range of the M-type stars 
\citep{dami18}, where the intrinsic locus 
bends towards a different direction, with a slower variation in 
the bluer B-V colors than in
the redder ones.

We find that assuming the  reddening relations
given in \citet{fior03} for $R_V$=3.1, the two shifted loci (green lines) do not   satisfactorily describe the data,
since only the M-type part of the intrinsic color relation 
can match our data. In contrast, assuming the anomalous wavelength reddening relations, $R_V$=5.0 by
 \citet{fior03}, the reddened loci (blue lines) can match our data along the entire range of spectral types.
We consider this as an evidence
 that the members of NGC\,6530 obey to the abnormal reddening law, rather than
to the standard one.

A further way to investigate the reddening law valid for the cluster members is to compare
the individual reddening ratios, derived spectroscopically as described in the previous section,
for several photometric bands and compare them with the analogous ratios expected
 for $R_V=$3.1 and $R_V=$5.0.
Also in this case, we used the sample of WTTSp with T$_{\rm eff}>4000$\,K. The individual reddening ratios as a function of the effective temperatures are shown in 
Fig.\,\ref{wttsratred}.
The observed spread is dominated by the individual errors rather than the intrinsic dispersion.
Nevertheless, despite the quite large dispersion, the mean values of the ratios  in the given photometric
color combinations, are in agreement, within the errors, with the abnormal reddening law R$_V=5.0$ rather  than
with the standard one, as shown in the bottom/right panel of Fig.\,\ref{wttsratred}.

Our data offer a further opportunity to  investigate the reddening law in the
region around NGC\,6530 by considering also the sample of giants.
As described in  Sect.\,\ref{giantsandms}, the sample of GES targets includes
a subsample of RC giants, i.e. stars in the core helium-burning phase \citep{gira99}.
 We expect most of them belong to the Bulge and are
located within a relatively small range of distances with respect to 
the other giants.  These objects 
have similar intrinsic astrophysical properties, 
such as temperature and luminosity, at a given metallicity and age. 
 These properties have been exploited in the past  to derive the reddening maps in several
regions  and also to derive the reddening law, as done in \citet{de-m14,de-m16}
 for the 30 Doradus Nebula. 
In fact, assuming similar magnitudes and colors,
their distribution in the CMD is a signature of the amount of interstellar reddening 
affecting them, while the slope of their distribution corresponds
 to the ratio R between absolute and selective extinction in the 
 specific color.

\begin{figure*}
 \centering
 \includegraphics[width=\textwidth]{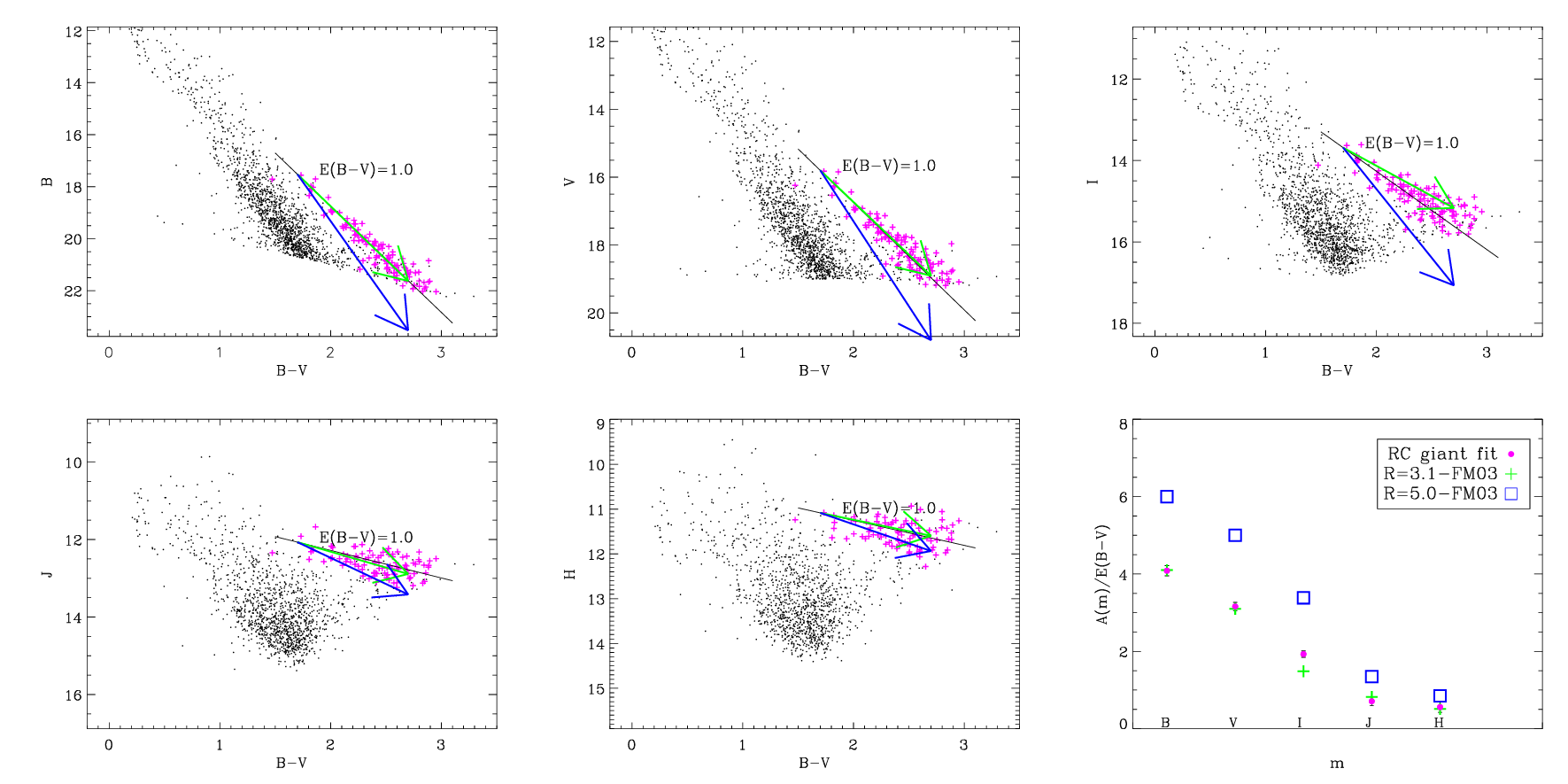} 
 \caption{First five panels: CMD of all GES targets (small black dots) of the BVIJH magnitudes 
 as a function of B-V.
 RC giants are indicated with magenta plus. Black solid line is the slope obtained by the fit,
green and blue arrows are the reddening vectors corresponding to the standard and abnormal
reddening laws, respectively.
Bottom/right panel: comparison of the  fitted slopes for RV giants in different photometric bands, with the literature reddening laws.}
\label{giantsredlawcmd}
 \end{figure*}

As in \citet{de-m14,de-m16},
we performed a linear fit in the CMD to the RC giants.
 We note that \citet{de-m14,de-m16} 
 identified a compact and well defined locus of the RC giants in the 30 Doradus CMD,
 that in our CMD  shows a larger spread,
 due to the larger apparent size of the Bulge compared to
  the LMC.
  However, also in our case, the slope
 of their distribution
 can be used to estimate the reddening law.
  The resulting slopes are overplotted in the  CMDs 
of Fig.\,\ref{giantsredlawcmd}, 
where the comparison with the \citet{fior03} relations
given for R$_V=3.1$ and 5.0 is also shown.
The uncertainty related to the range of distances in the Bulge cannot 
affect significantly our results.

\begin{figure*}
 \centering
 \includegraphics[width=\textwidth]{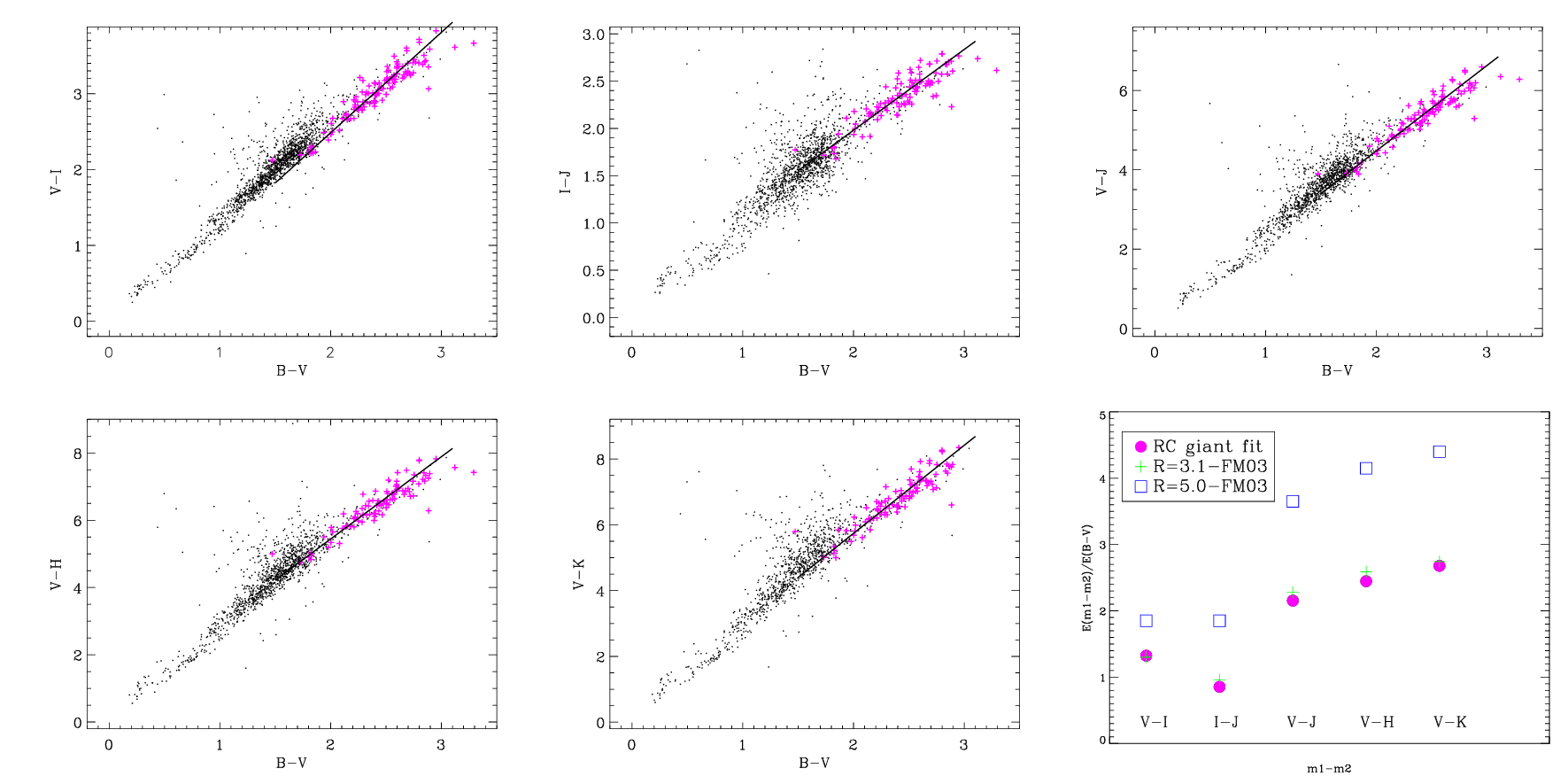} 
 \caption{First five panels: color-color diagrams of all GES targets.
 RC giants are indicated with magenta plus symbols. Black solid line indicates
 the slope obtained from the fit.
Bottom/right panel: 
comparison of the obtained slopes with the literature reddening is shown.}
\label{giantsredlaw}
 \end{figure*}

We performed the analogous fit on the color-color diagrams and we obtained the results shown in 
Fig.\,\ref{giantsredlaw}. We note that the color-color
 diagram slope fitting is independent of the distance and 
therefore  the results are more robust than those obtained from the CMD. However, the  fitting results, both in the CMD and in the color-magnitude diagram suggest that
unlike the cluster members, 
RC giants do not follow an abnormal reddening law, but the standard one.

  \begin{figure*}
 \centering
 \includegraphics[width=\textwidth]{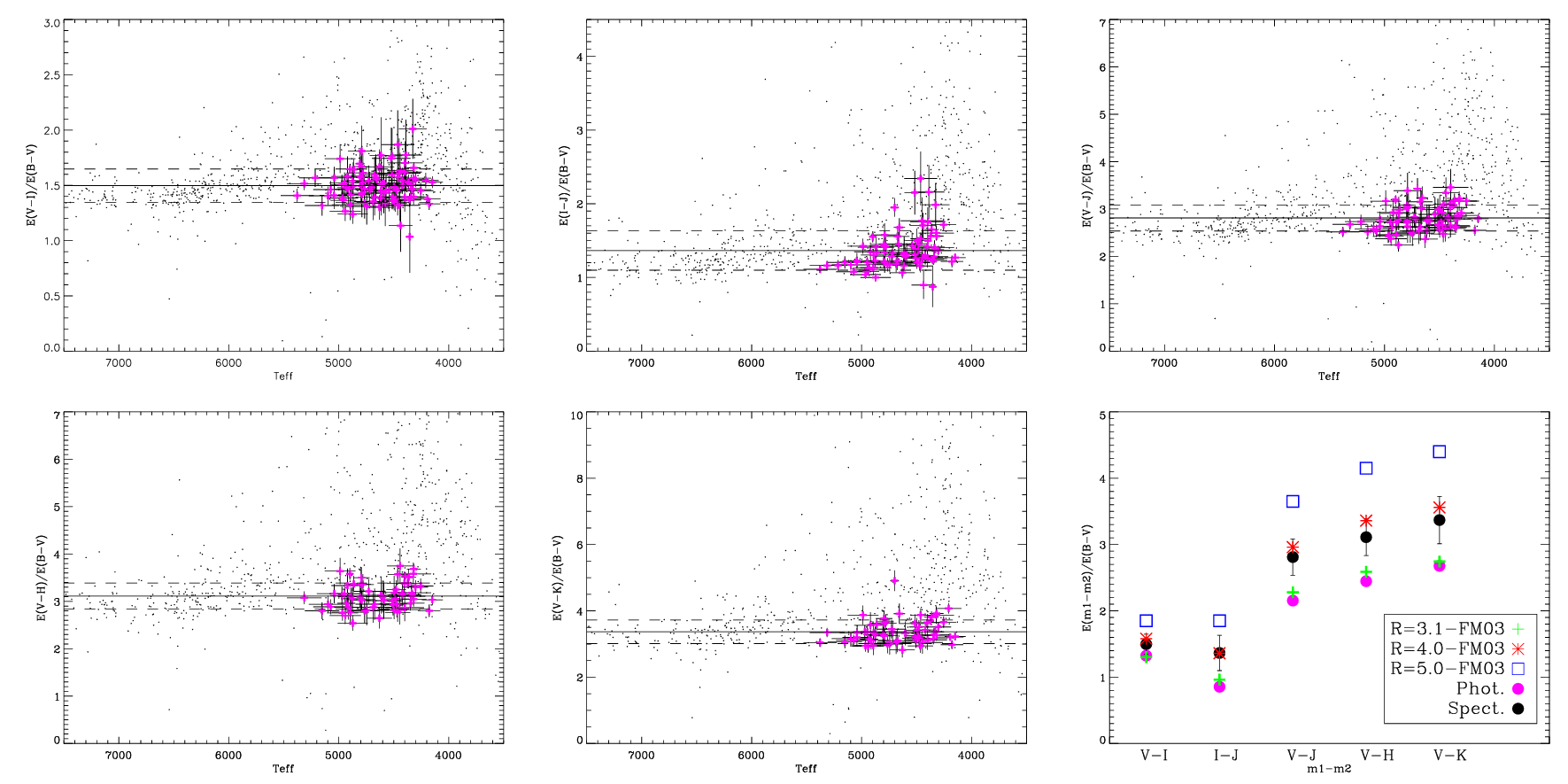} 
 \caption{First five panels: 
 reddening ratios in several colors over E(B-V) for all GES targets.
 RC giants are indicated with magenta plus symbols. Black (dashed) solid line is the mean (standard deviation),
obtained for RC giants. 
Bottom/right panel: comparison of the  slopes obtained from the photometric and spectroscopic data
 with those given in the literature for different reddening laws.}
\label{giantsratred}
 \end{figure*}

As we have done for the cluster members, we also used  the individual reddening derived spectroscopically
  and computed the ratios 
between the reddening in several colors and E(B-V). In this case, the statistical errors on  
the effective temperatures as well as possible systematic errors on 
 reddening due to the sample selection
(mis-classification of the luminosity class) or to the color-temperature relations, contribute to 
enhance the observed spread in the reddening ratios.
In order to reduce such dispersion, we selected only the RC giants with errors in 
 reddening ratios
smaller than 0.4 mag. Figure \ref{giantsratred} shows such ratios as a function of the effective
temperatures.  The comparison of the mean ratios with the analogous ratios found from the color-magnitude diagram and that
with the values given in \citet{fior03} are also shown.

The spectroscopic data suggest that the reddening law  of the RC giants is marginally consistent
with the standard reddening law, unlike the photometric ones. Instead, spectroscopic results suggest
an intermediate reddening law, compatible with an intermediate R$_V$ value, like R$_V$=4.0.
The background giants should experience a combination of extinction due to both the large dust grains in the Lagoon Nebula (with an abnormal reddening law) and the background interstellar extinction (with a standard reddening law). For this reason, 
we expect the reddening law measured for the background stars to be somewhere between the Rv=3.1 and Rv=5.0 reddening laws, as suggested by our spectroscopic results.

%The origin of this discrepancy might be due to the color-temperature relations adopted %for giants 
%\citep{bess88}. 
%However, the photometric results are affected by larger uncertainties than the spectroscopic ones, 

In conclusion we find that while towards cluster members an anomalous 
reddening law is more appropriate to derive the stellar  absorption, around background
objects, such as the RC giants, the reddening law is, instead, between standard
and abnormal.

The indication of an anomalous reddening law around cluster members,
 obtained in two independent ways,
supports the hypothesis of the prevalence of large dust grains with respect to small grains.
This can be explained with the evaporation of small grains by the radiation of O-B type stars,
or with the grain growth in the circumstellar disk \citep{toth08}, in agreement with recent planet
formation theories \citep{mord12,gonz17}.

\section{Spatial distributions and nebula structure}
The open cluster NGC\,6530 and the surrounding nebula have been described in the literature
 as having a complex spatial structure \citep[e.g.][]{toth08}. As already mentioned in the previous sections, our data provides us with 
 fundamental stellar parameters, such as effective temperatures,
gravities and reddenings not only for the cluster population but also for the field stars
falling in the same field.
 In the following we show how these parameters can be used to derive hints on the 
thickness of the nebula. 

Individual reddenings found for the different populations of stars give us also indications
about their expected distance and relative position, since our ability to reach objects at very large distance 
is strongly related to the  transparency of the material between us and them.
 Therefore, the properties of these objects
can be used as tracers of the interstellar material.

  In this section we study the spatial distribution of the three main populations selected in this work,
i.e. cluster members, MS and giant field stars, with the aim of connecting their location on the sky
with the properties of the nebula surrounding the cluster NGC\,6530.

 \begin{figure}
 \centering
 \includegraphics[width=8cm]{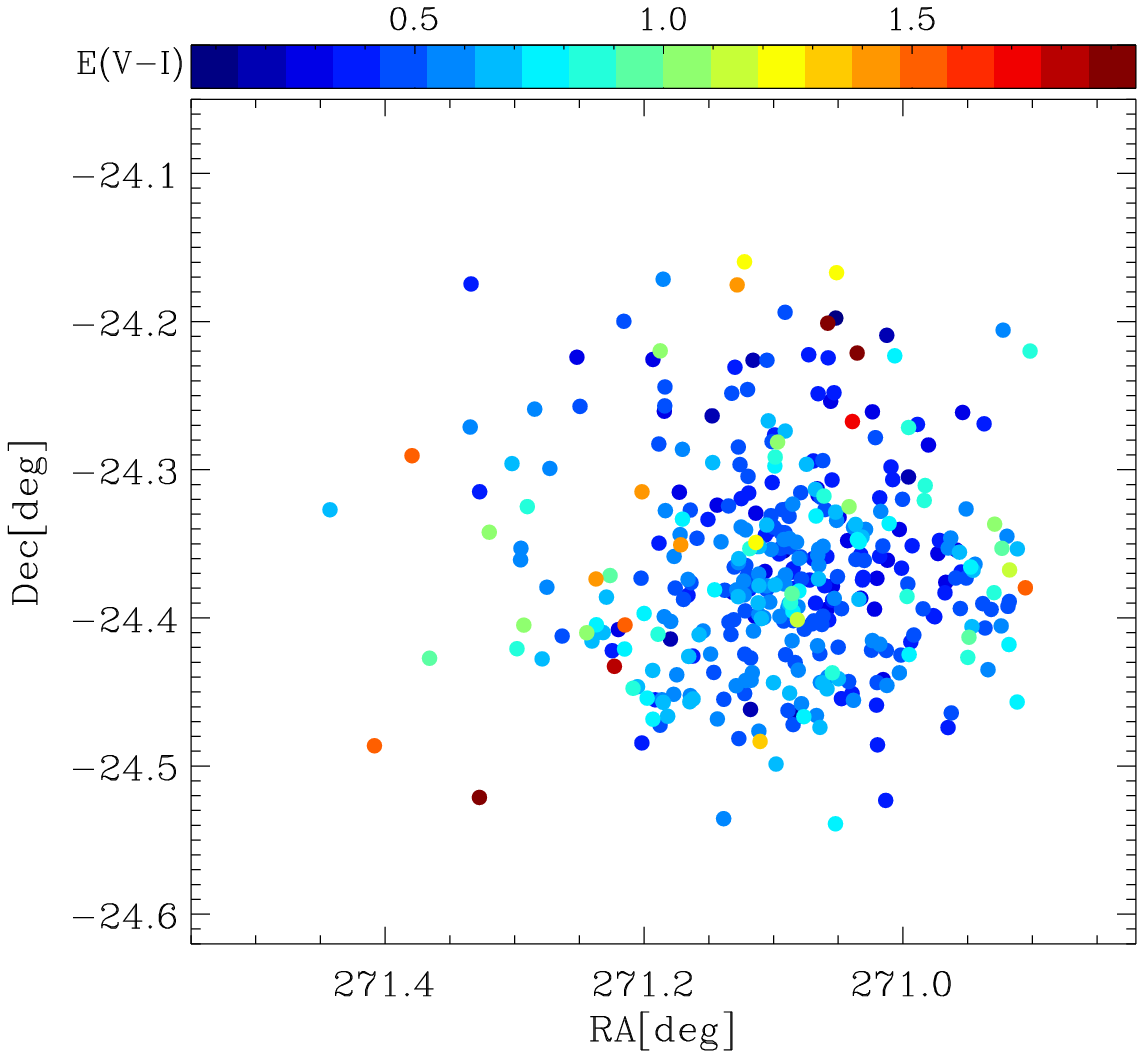} 
 \includegraphics[width=8cm]{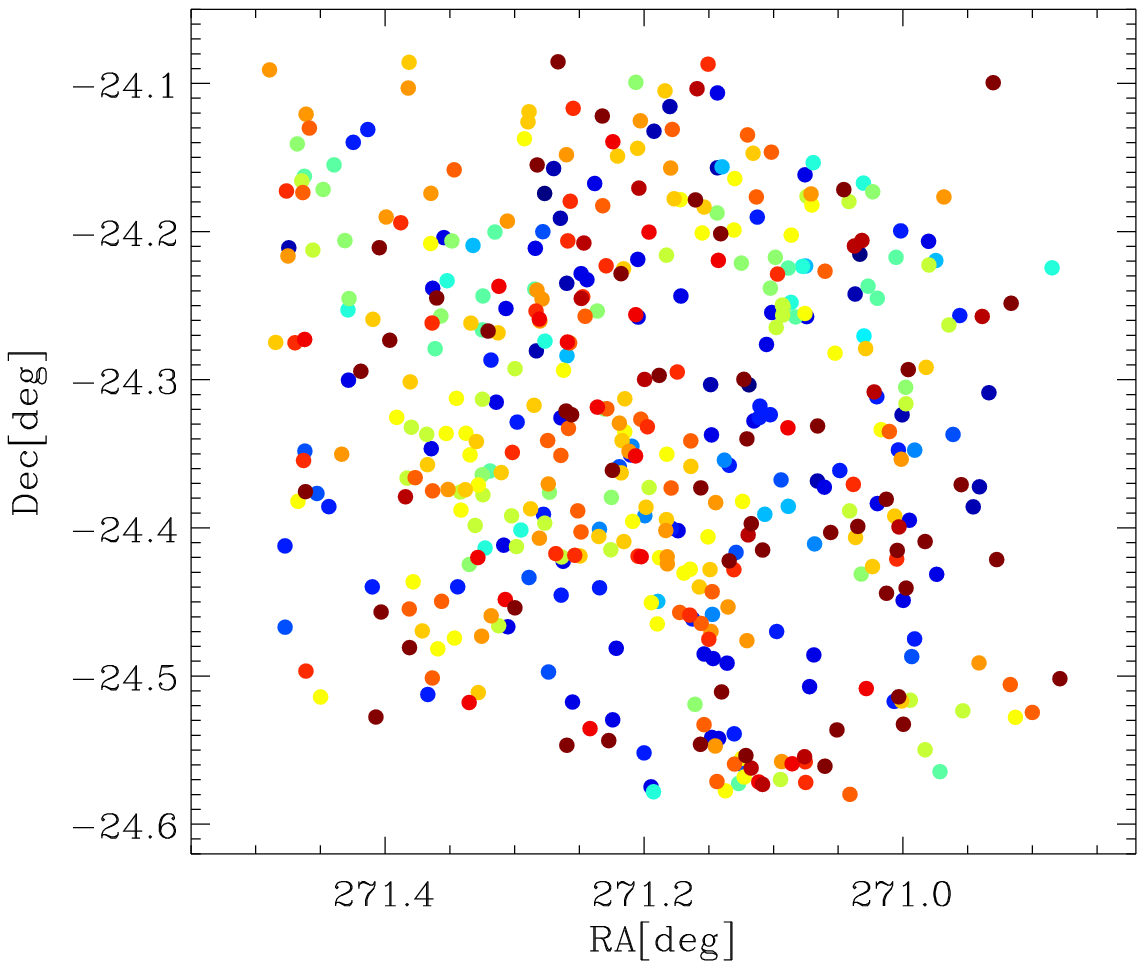}
 \includegraphics[width=8cm]{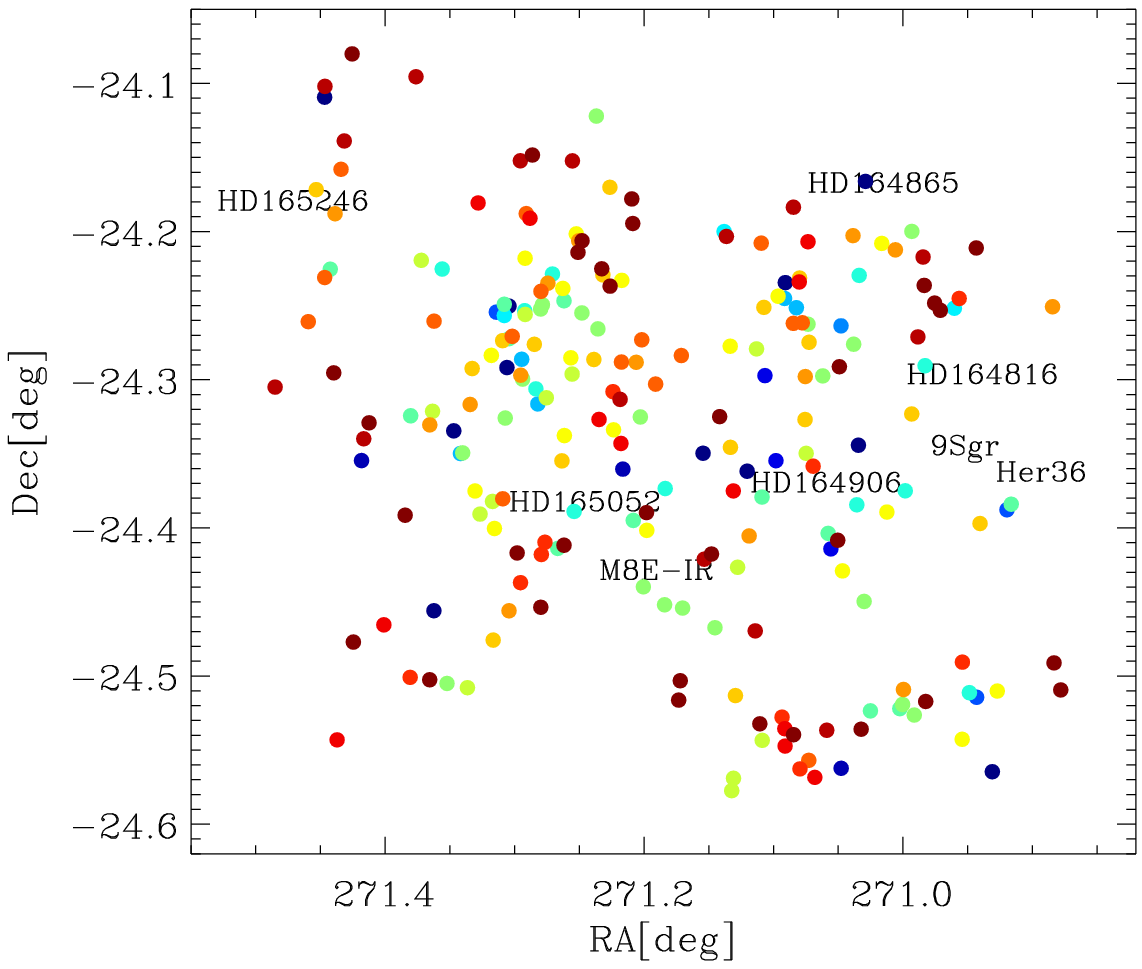}
 \caption{Spatial distributions  of cluster members (upper panel),
 candidate MS stars (middle panel) and giant stars (bottom panel),
 for which the reddening has been derived. Symbol colors are drawn as a function
of E(V-I) as defined in the  color bar. For a better readability,
the positions of massive stars are shown  in the bottom panel only.}
\label{popredmap}
 \end{figure}

Figure\,\ref{popredmap}  shows the spatial distribution of 
cluster members, MS and  field stars for which the interstellar reddenings have been derived, i.e.
the GES targets  with
 optical WFI photometry \citep{pris05}.  
The comparison shows that the cluster is dominated by objects with
 reddening significantly different from that affecting MS stars and giants.
Cluster members are mainly concentrated in the region delimited by
 the subclumps found by \citet{kuhn14}, where most of the OB-type stars are found while
 the populations of MS and giants are more randomly distributed.
  
% with a quite elongated distribution along the direction defined by the stars HD164614 and M8E-IR.

%We also note a spatial correlation of cluster members with the E(V-I) values,
%with the objects with 
%0.2$<$E(V-I)$<$0.3 being mainly located around the cluster center and %
%	in the NW region of the cluster,
%members with  0.3$<$E(V-I)$<$0.4
%	 are quite concentrated around the cluster center with several objects
% dispersed on the Weastern region, members with 0.4$<$E(V-I)$<$0.5 are more dispersed but lie mainly in the
%weastern and in the southern regions, 
%while members with  0.5$<$E(V-I)$<$0.7 are mainly located in the southern region.
%Finally, the few members with 0.7$<$E(V-I)$<$0.8 are mainly distributed along
% the cluster direction, i.e. between the stars HD164614 and M8E-IR.
The spatial distribution of   MS stars
 and  giants  form a patchy pattern with subregions completely devoid of  stars.
%For example 
%the region around Herschel\,36,
%the region between HD164865 and HD164816,
%the two regions on  east and on  weast of the cluster center
%and, finally, the region on weast of HD165052
%are almost empty of field stars. 
 We note, however, that the few MS field stars 
affected by very small reddening, expected to be foreground objects,
 are quite uniformly distributed on the field, while the peculiar 
pattern is mainly drawn by the stars affected by higher reddening,
 expected to be located at distances similar  to
 the cluster or further away.
 We interpret this latter peculiar spatial distribution
 as evidence of the non-uniform  transparency of the surrounding Lagoon
 Nebula,  connected to its thickness.

 In particular, there are two regions,
 roughly centered at (271.15, -24.27) and (271.07, -24.5),
that are  almost empty of background  stars.
We suggest these are the darkest and thickest  parts of the nebula. 
where background stars are too extincted to be detected with
our observational limits. In contrast, 
most  of the reddened field stars are  found outside these regions, that
correspond to less extincted  areas of the Nebula, where 
the thickness is instead lower.

The spatial distribution of  giants is characterized by a lower stellar density
but a pattern similar to that found for MS background stars 
since the subregions devoid of (filled with) MS stars correspond to those
 devoid of (filled with) giants. 

This confirms our hypothesis on the nebula structure.
In fact,  RC giants are very distant objects and are expected to be homogeneously distributed  across the sky
 since the cluster NGC\,6530 is located  close to the direction of the Galactic center.
Our results indicate for RC giants increasing reddening values
(with respect to cluster members) a fact that  is  consistent 
with the hypothesis of very distant objects with an apparent 
inhomogeneous spatial distribution,
that is correlated with the interstellar material transparency. 
This confirms that the regions devoid of  giants,
 are likely very opaque areas where the Lagoon Nebula dust 
prevents detection of objects behind it.

The anticorrelated spatial distributions of cluster members and field stars, and the consequent evidence
of a different transparency of the nebula around these different
stellar populations, is consistent with 
the different reddening laws found in Sect.\,\ref{redlawsect} for cluster members and for giant stars. In fact,
the reddening law is strongly dependent on the grain size distribution of the interstellar material \citep{math81}. In the direction of the cluster center, small dust grains of the background interstellar medium are likely 
not detected due to the thickness of the nebula. In addition, due to the evaporation of small  grains by radiation from massive stars
and the presence of the large number of YSOs with circumstellar disks,
the grain size distribution of the nebula 
is expected to be dominated by larger-than-average grains. 
 In contrast, in the regions where the nebula is not opaque,
the  reddening is  dominated by the interstellar material properties up to very large distances,
and therefore by a standard reddening law.

As shown in Sect.\,\ref{redlawsect}, the  reddening law appropriate to NGC\,6530
cluster members is $R_V=5.0$. Using this relation and 
the individual reddening values E(V-I)
obtained as described in Sect.\,\ref{intred},
we derived individual  absorption values $A_V$  for 
the samples including
%RESTORE,'../SAVEFILES/intrinsic_colors_pisa.save',/v
%clII =where(membership.fin_mem Eq 1 and (membership.disk EQ 1 or membership.accretor EQ 1)  and $%
	%	 finite(data.evi))
%clIII=where(membership.fin_mem EQ 1 and membership.disk NE 1  and membership.accretor NE 1 and $	
	%	finite(data.evi));  
%allmass=where(membership.fin_mem EQ 1 and finite(data.evi));
%help,clii,clIII,allmass
%CLII            LONG      = Array[240]
%CLIII           LONG      = Array[147]
%ALLMASS         LONG      = Array[393]
the \memwttsage WTTSp members 
 and the \memcttsage CTTSe  members with WFI V and I photometry.
For these samples
we also computed stellar luminosities, masses and ages 
using the recent PISA stellar evolutionary  models for solar metallicity 
 and the cluster distance \clustdist\,pc that is the weighted mean distance 
found in Sect.\,\ref{intred}. 

 \begin{figure*}
 \centering
 \includegraphics[width=\textwidth]{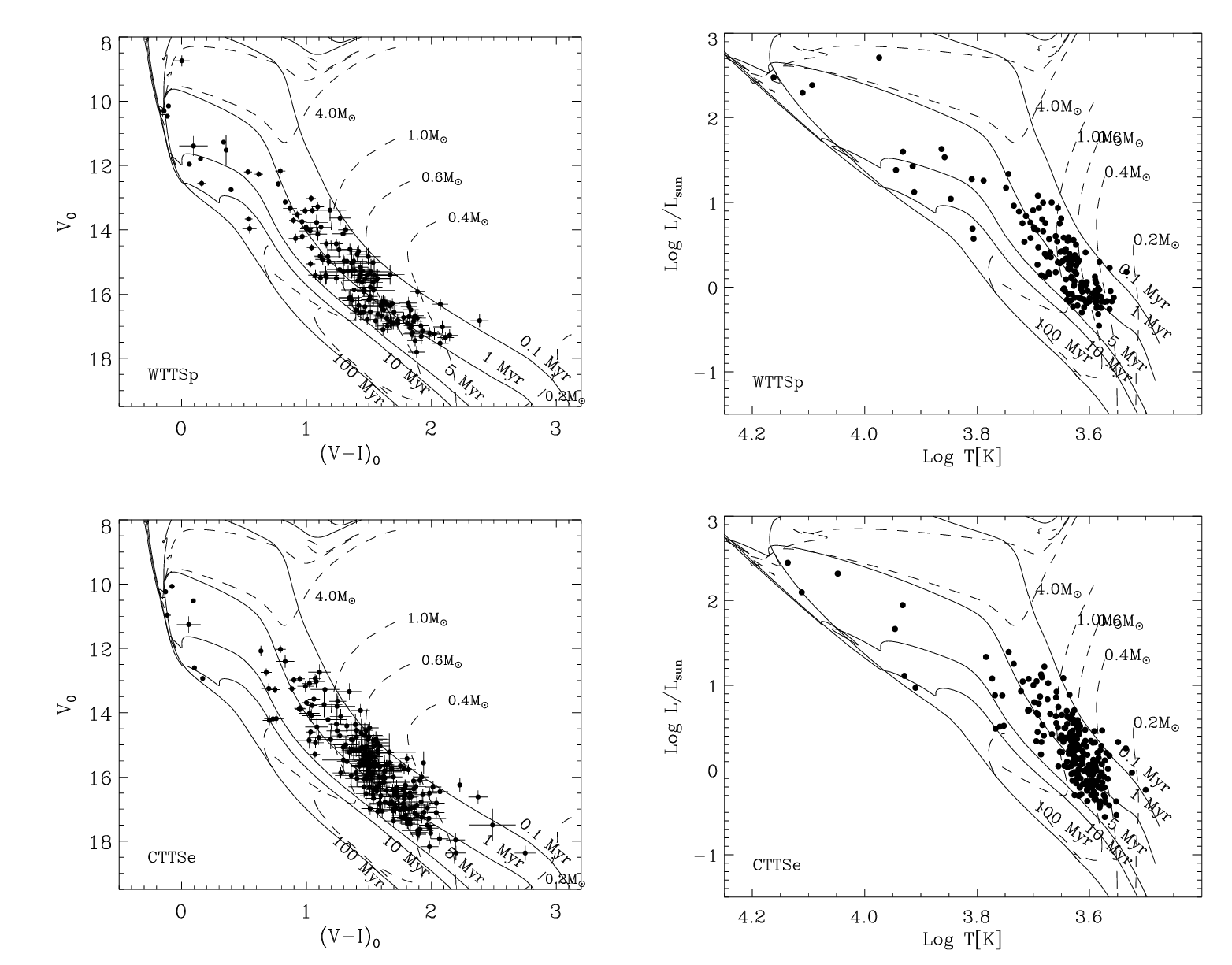} 
 \caption{V$_0$ vs. (V-I)$_0$ (left panels)
and the HR (right panels) diagrams for members classified as 
WTTSp (upper panels) and CTTSe (bottom panels).
Dashed and solid lines are the PISA evolutionary tracks and isochrones 
at masses and ages indicated in each panel, and at the cluster distance. }
\label{diagramspisa}
 \end{figure*}
Figure\,\ref{diagramspisa} shows the intrinsic V$_0$ vs. (V-I)$_0$
and the HR diagrams for  the \memwttsage WTTSp  and the \memcttsage
 CTTSe  members, compared with the adopted tracks and
isochrones.
Isochronal ages were assigned by the bilinear interpolation
of the adopted isochrones to the position of cluster members in the 
V$_0$ vs. (V-I)$_0$ diagram. Age values are given in Table\,\ref{tablemem2}.

\begin{figure}
 \centering
 \includegraphics[width=9cm]{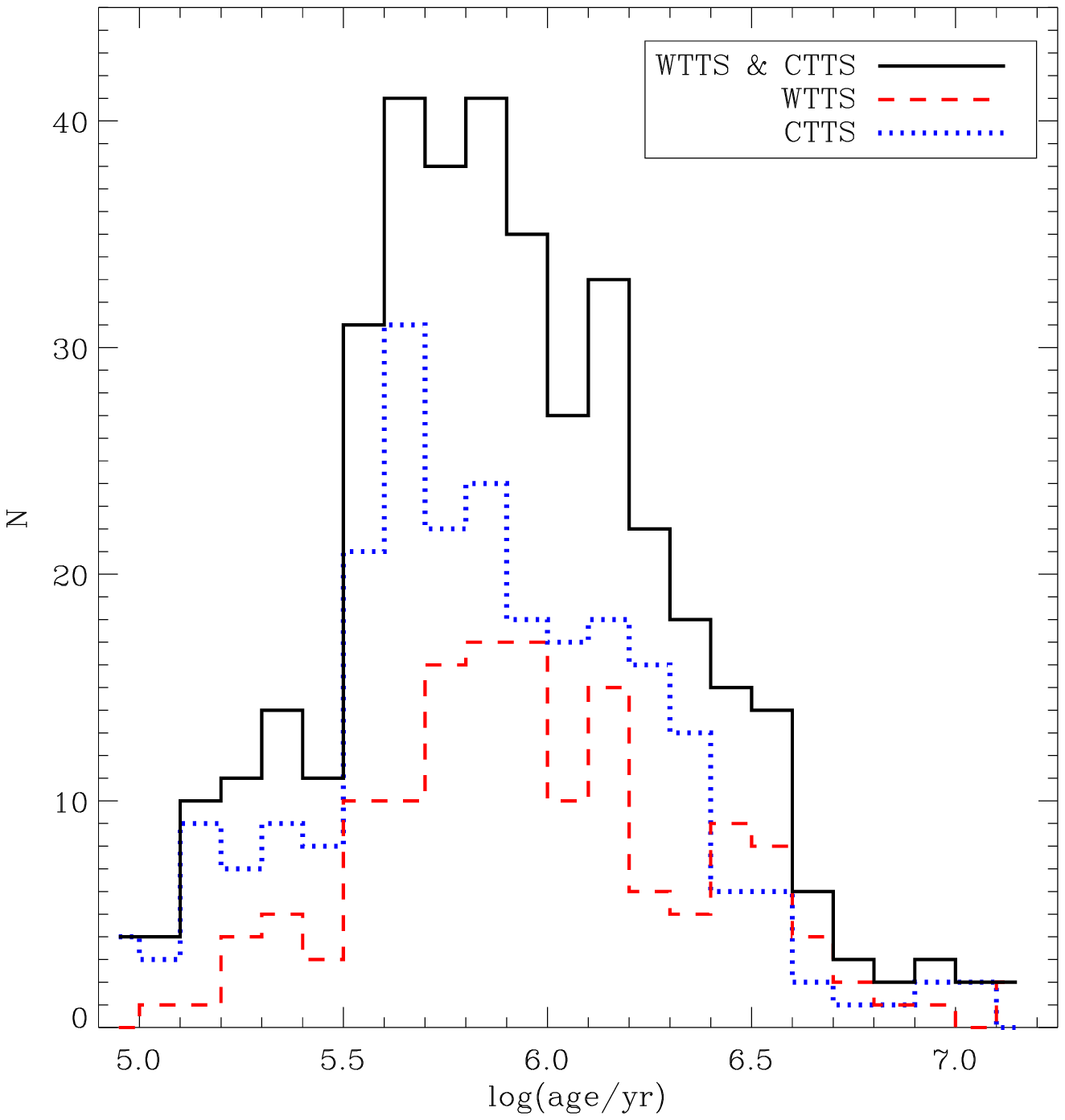} 
 \caption{Isochronal age distribution of all members, WTTSp  and CTTSe, 
 drawn as solid/black, dashed/red and dotted/blue histograms,
 respectively. }
\label{agehist}
 \end{figure}

Figure\,\ref{agehist} shows the histograms of isochronal stellar ages 
derived for all members, WTTSp and CTTSe.
The distributions are approximately log normal, and therefore we 
 quote means and dispersions in dex rather than linear units.
 
We find that the hotter stars are, or look, older than  cool stars, 
this being  a common problem (yet unsolved) when looking at star forming regions.
For this reason we did not include hotter stars  in the  calculation
of the statistical parameters of the sample. 
Cluster members with T$_{\rm eff}<5500$K  have masses in the range 
[0.24-2.80]\,M$_\odot$ and ages between
0.1 and 5\,Myr. % output diagrams_pisa_cool.pro
%There are only 10 members  with age older than 5\,Myr.
%The mean mass for WTTS or Class\,III (CTTS or Class\,II) stars is 
%1.0\,M$_\odot$ (0.8\,$_\odot$) with a standard
%deviation of 0.6\,$_\odot$ (0.4\,$_\odot$). 
The mean log age for all members is 5.84 (units of years) with a dispersion
of 0.36 dex. If we consider WTTSp and CTTSe separately, the mean log age is 
% see diagrams_pisa_cool.pro
  5.92 for WTTSp and 5.81 for CTTSe with dispersions of 0.35 and 0.37 dex,
  respectively.
 The two latter distributions are marginally different as confirmed by  
 the Kolmogorow-Smirnov test (KS test), that returns a probability 
 of 0.016 that the two 
 distributions come from the same parent sample. 
  As expected, the number of CTTSe with younger
 ages is larger than that of WTTSp but the two dispersions are very similar.

To establish if the age dispersion is significant, or dominated by the total
errors on stellar ages, we   quantified the errors on the isochronal 
stellar ages  by performing Monte Carlo simulations, as described 
in the following section.
\subsection{Error estimates: Monte Carlo simulations}
Errors on ages  depend on several observational uncertainties, affecting 
our ability to accurately position each star in the V$_0$ vs. (V-I)$_0$ 
diagram, and also on the stellar variability which causes the stars move across the diagram.
In addition, they depend on the model accuracy. 

Errors on (V-I)$_0$  depend on the errors in the spectroscopic effective 
temperature, through the color-T$_{\rm teff}$ relation adopted in our models.
We note that the $\sigma({\rm T}_{\rm eff})$ values adopted here were obtained
by considering 
  the statistical errors in  T$_{\rm eff}$ released by the Gaia-ESO survey,
and  the systematic errors due to the
calibrations, given in \citet{dami14}. These latter amount to
 109, 73 and 50\,K for stars with 
T$_{\rm teff}>5500 K$, 4300$<$T$_{\rm teff}/[K]>5500 K$ and
 T$_{\rm teff}>4300 K$,
respectively. 

Error bars of V$_0$ depend on 
the   errors in the observed magnitude V and on the error 
in the absorption 
A$_{\rm V}$. This latter is derived from  E(V-I),
and therefore depends on  the reddening law,
the  errors in the observed 
color V-I and the errors on (V-I)$_0$.

Global uncertainties on  V and V-I include the photometric errors
and the effects of photometric variability, in case of stars with 
variable extinction, accretion bursts, and  cool or hot spots 
 \citep[e.g.][]{gull98,bara09,cody14,stau16}. 
All these kinds of variability
are expected to affect our ability
to derive stellar ages  for  CTTSe, while stellar ages of WTTSp can be affected only by 
variability due to cool spots. Nevertheless, the presence of a 
 comparable dispersion  in the stellar age distributions of both
WTTSp and CTTSe 
 suggests that the observed spread is not mainly determined by  photometric
 variabilty due to accretion, hot spots,  
 or scattering from reflection nebulae 
\citep[e.g.][]{gros03,luhm07}.  
To evaluate the uncertainty due to photometric variability, we used the
results by \citet{hend12}, who found 
 0.02$\leq \Delta {\rm I} \leq $  0.2, and 
 we estimated the variability in the V band, using the dI/dV values
found by \citet{herb94} for a sample of WTTSp.

For each star, we generated three sets of 1000 values,
%monte_carlo_ages.pro
normally distributed, with dispersions equal to 
the  errors in
 T$_{\rm eff}$, V and I of that star.
We computed a further normal distribution of 1000  $\Delta {\rm I}$ values
with mean $<\Delta {\rm I}>$=0.11 and $\sigma(\Delta {\rm I})$=0.03 that
roughly corresponds to a distribution of values between 0.02 and 0.2, as
found in \citet{hend12} and finally 
we computed a normal distribution of 1000 dI/dV values with 
mean $<$dI/dV$>$=0.67 and $\sigma$(dI/dV)=0.13, these latter parameters
being derived from the \citet{herb94} values.

We added these five sets of random errors to the observed values V$_0$ and (V-I)$_0$
and then computed 1000 values of stellar ages for each star.
The statistical error of the  age of each star is the standard deviation
of these 1000 values. The typical logarithmic standard deviation (mean value)
that we found is 0.09 dex % see read_monte_carlo output.
that we assume as representative mean statistical
uncertainty in the isochronal ages.

In our  simulations  we did not take into account 
the error on  the cluster distance,
used to locate the isochrones in the V$_0$ vs. (V-I)$_0$.
At the distance of NGC\,6530,  this systematic uncertainty affects 
the position of the isochrones  only in the vertical direction of the diagram
 V$_0$ vs. (V-I)$_0$
and would imply the same and
 constant systematic error on all stellar ages, affecting the accuracy 
 but not the precision, and hence the dispersion in which we are interested.   

We did not consider the uncertainty introduced
by the adopted models because 
 \citet{regg11} showed that   
by adopting three different
sets of isochrones, even though the amount of the age spread is slightly 
model dependent,
it is significantly large, regardless of which family of models is
adopted.

Finally, our calculation does not include the contribution by unresolved binary
companions, since this is not a random error but a systematic bias towards
 higher luminosities (L), and therefore  younger ages.
The  additional contribution to the luminosity of a single star 
due to a companion can be described  using the  $\Delta$ log L
 distribution   given  in  \citet{hart01}, ranging
 from 0.05 (for the minimum mass companion) to 0.3 (for the
equal mass companion). 

For estimating the uncertainty due to unresolved binaries, 
we performed a Monte Carlo simulation, 
by generating a  distribution of 
1000 $\Delta$ log L values  between 0.05 and 0.3 as in \citet{hart01}.
Then, we generated a coeval  sample of 1\,Myr old artificial stars 
with (V-I)$_0$ equal to those of 
our members and V magnitudes from the 1\,Myr isochrone.

To the V magnitude of each artificial star, we added  the
  dV values consistent with
the $\Delta$ log L binary distribution and computed the corresponding
 stellar ages.
As for the statistical error, we computed the standard deviation of the isochronal
ages and found that the typical uncertainty due to unresolved binaries 
is 0.10 dex 
%monte_carlo_binaries.pro
%see read_monte_carlo_binaries.pro
with a shift $\Delta$ log t=-0.21 dex 
with respect to the starting stellar age of 1\,Myr, in agreement with
the results found in \citet{hart01}.

By combining the statistical age error equal to
0.09 dex, derived above,  and the systematic age spread equal to 
0.10 dex, due 
to unresolved binaries,  we found that the typical total uncertainty
 in the isochronal
 ages is 0.13 dex, that is significantly lower than the observed spread 
equal to 0.36 dex.
%print,sqrt(0.0929293^2+0.0962877^2)
%     0.133818  
   
 We conclude that observed age dispersion, obtained by considering 
 all members and    the   WTTSp and CTTSe samples separately,
  cannot be accounted for by observational
  uncertainties and is evidence of a small but real age spread. 
 This would imply that star formation in NGC\,6530 
 occured within few Myrs, in agreement with what was found
 also in other SFRs,
 such as the Orion Nebula Cluster \citep[e.g.][]{regg11} and
  NGC2264 \citep{venu18}.
 
%This allows us to discard the hypotheis that  our estimation of stellar age
% errors are underestimated and hence we are confident that
%  the observed age spread is not influenced by the observational uncertainties.
  
%Figures\,\ref{cummass} and \ref{cumage} show,
%Figure\,\ref{cumage} show the cumulative age  distributions,
%for WTTS and CTTS members. 
%The mass distributions of the two samples  indicate
%that the CTTS star sample includes an excess of YSOs with mass around 0.8\,M, while in the WTTS star sample,
%the masses are quite uniformly distributed. The  two-sample  (K-S) test 
% applied to the two samples indicates a probability equal to 7.4e-5 that  the mass distributions
%  are extracted from the same parent distribution and this confirms that
%the two mass   distribution are significanly different.

\begin{figure*}
 \centering
 \includegraphics[width=\textwidth]{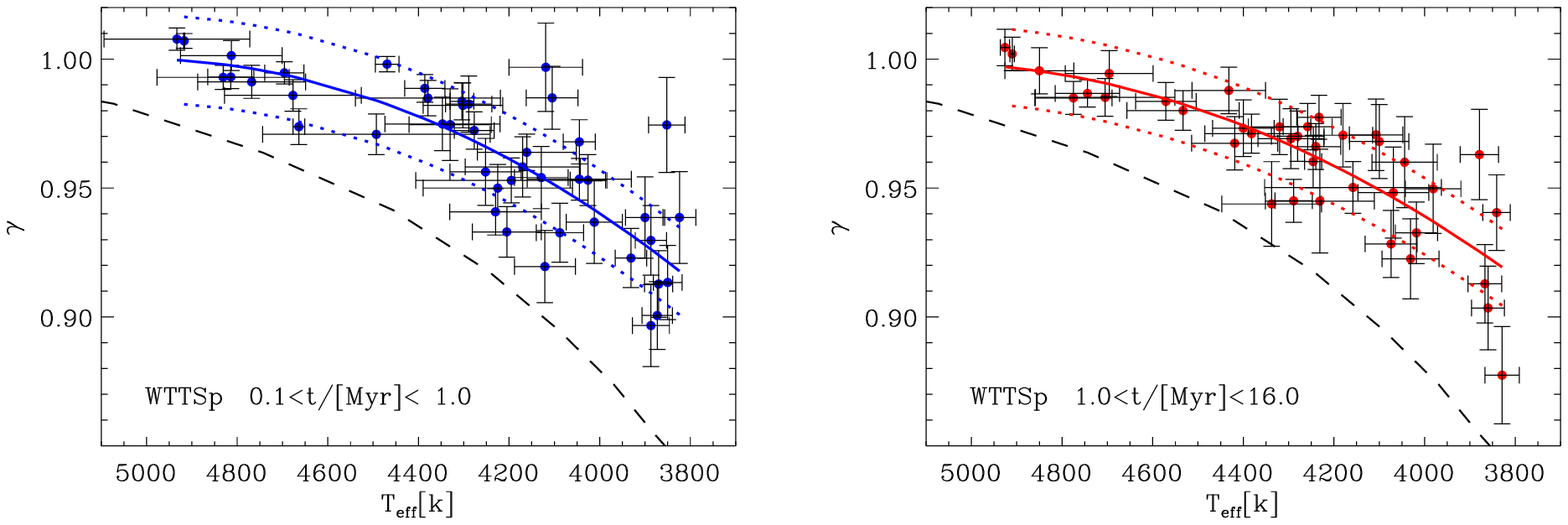} 
 \includegraphics[width=\textwidth]{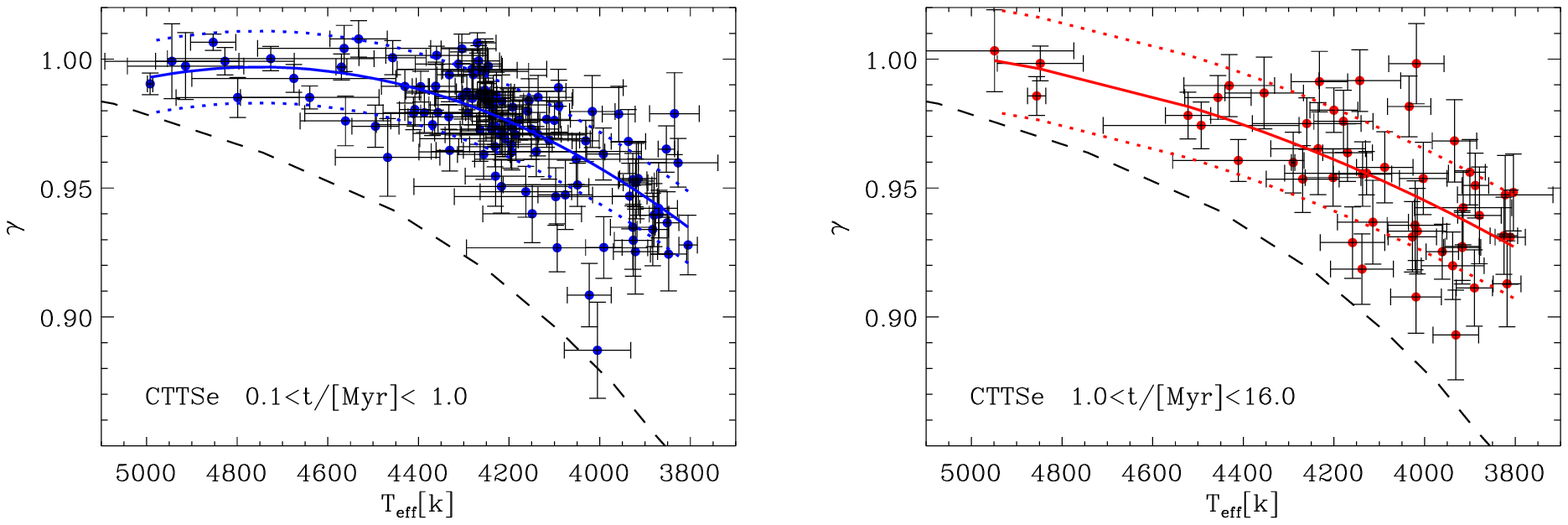} 
 \caption{$\gamma$ index as a function of T$_{\rm eff}$ for WTTSp (upper panels)
  and CTTSe (lower panelss)
 cluster members
 split into two different age ranges. Solid lines are the best 
fits obtained from a second order polynomial fit of the data while the dashed 
line is the reference locus
of MS stars obtained in \citet{dami14}. Dotted lines indicate 
the polynomial fit at 3$\sigma$. }
\label{gammaages}
 \end{figure*}
\begin{figure}
 \centering
 \includegraphics[width=9cm]{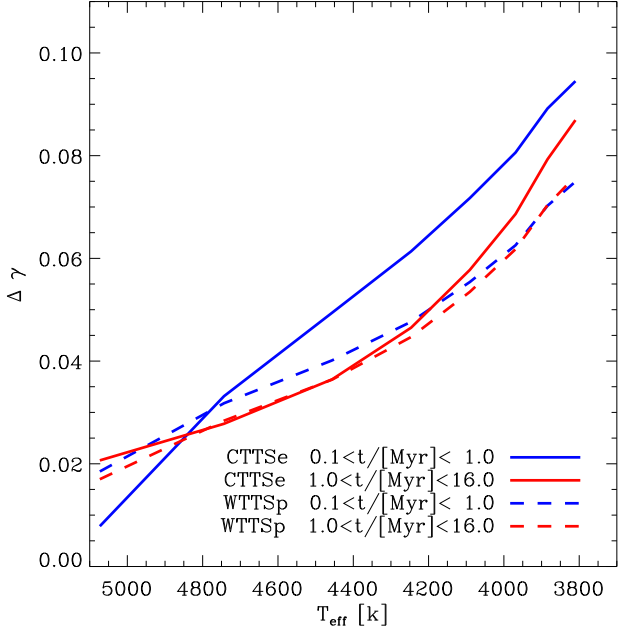} 
 \caption{$\Delta \gamma$ of CTTSe and WTTSp 
 (solid and dashed lines, respectively) as a function of 
 T$_{\rm eff}$ obtained from the difference between the best fit of the 
youngest and oldest stars  (blue and red lines, respectively)
and the reference MS locus.}
\label{gammaages2}
 \end{figure}
 
\begin{figure*}
 \centering
 \includegraphics[width=\textwidth]{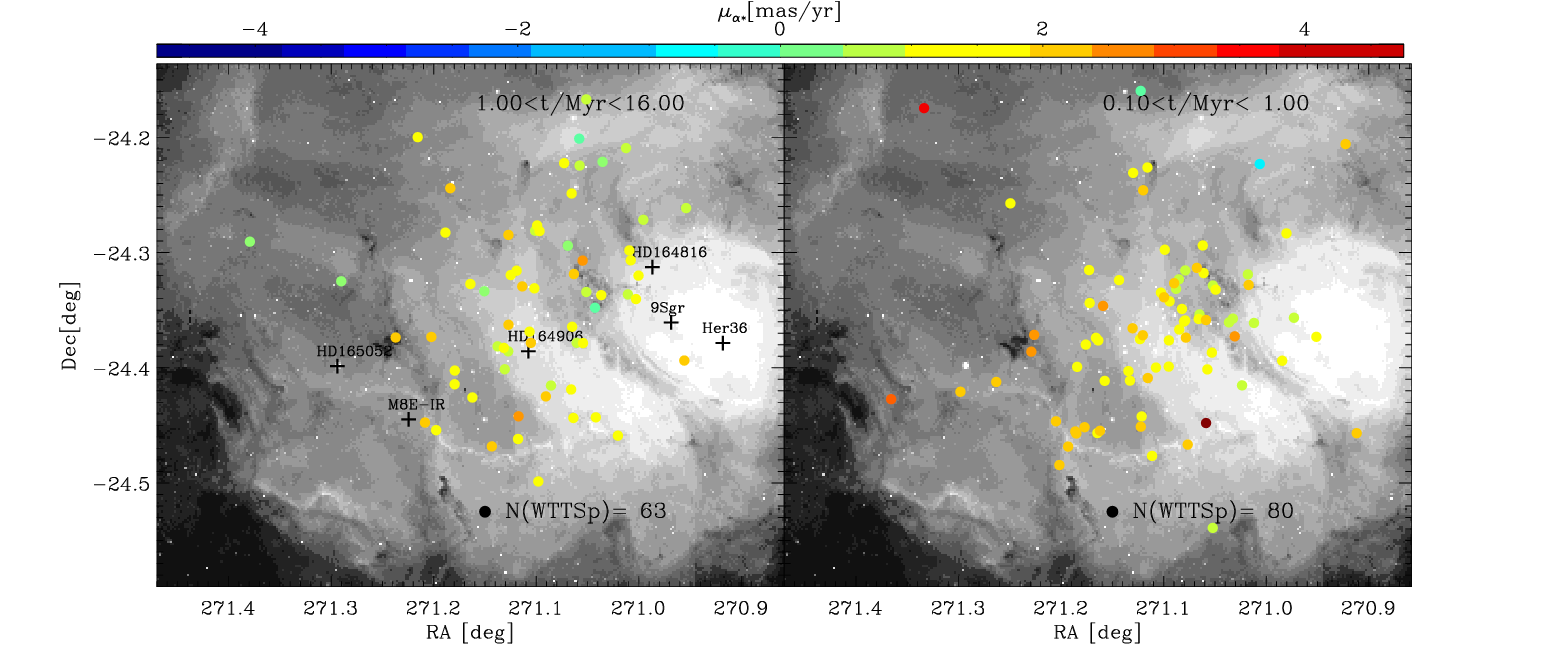}
 \caption{Spatial distributions of WTTSp
cluster  members split into two different age bins,
superimposed on a VPHAS+ H$\alpha$ image \citep{drew14}. Color codes
indicate the \emph{Gaia} DR2 proper motions in the right ascension.}
\label{memcliiiagebinmap}
 \end{figure*}

\begin{figure*}
 \centering
 \includegraphics[width=\textwidth]{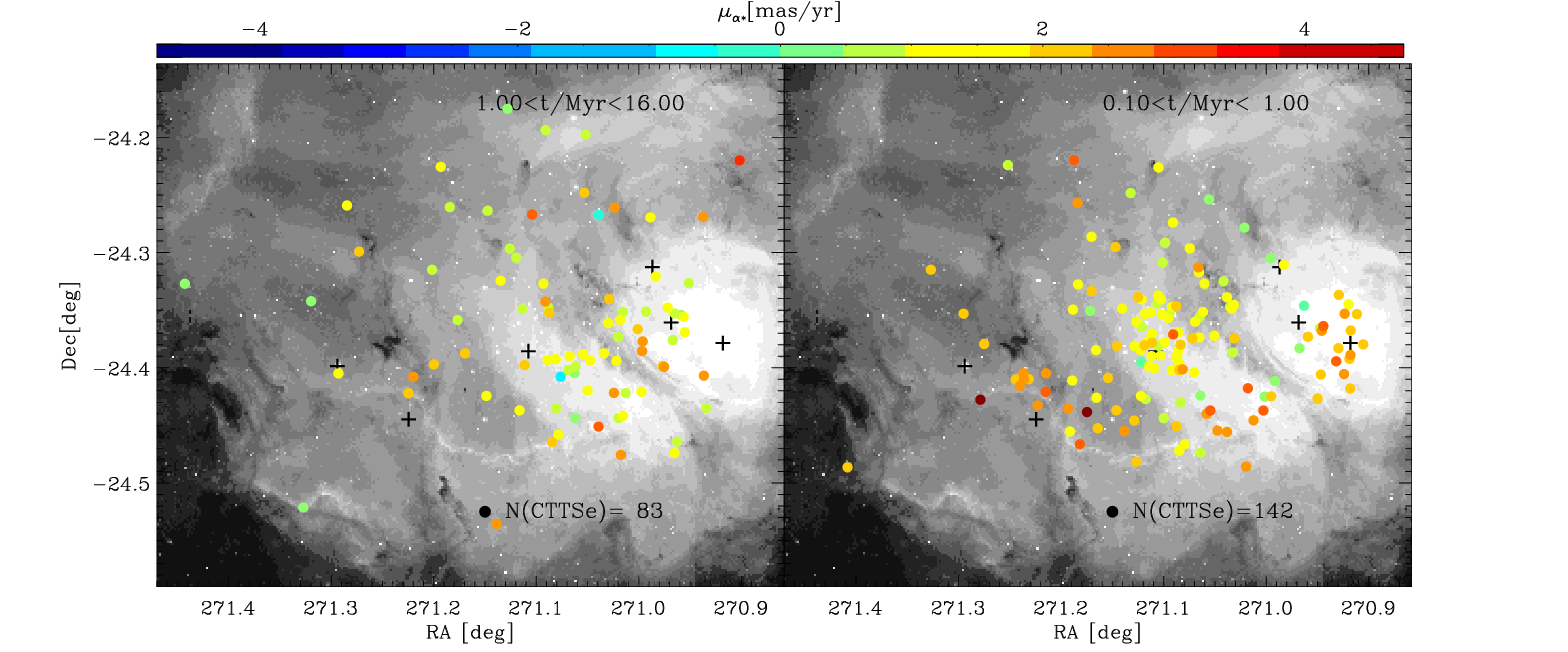}
 \caption{Spatial distributions of  CTTSe
cluster  members split into two different age bins,
 superimposed on a VPHAS+ H$\alpha$ image \citep{drew14}.
Color codes
        indicate the \emph{Gaia} DR2 proper motions in the right ascension.}
\label{memcliiagebinmap}
 \end{figure*}
 
\subsection{Gravity spread}
To evaluate if the observed age spread is supported by other
observational evidences,
we used a further indicator, i.e.
the gravity-sensitive $\gamma$ index, that changes with stellar ages
\citep{dami14}, 
expecially for very young stars. 
In fact, the aforementioned authors found that the gravity of intermediate and 
low mass stars  for the
  Chamaeleon\,I (1-3\,Myr) cluster, 
is lower than that obtained for  the $\gamma$ Vel cluster (5-10\,Myr). 
As a consequence the $\gamma$ index decreases with ages down to the
typical values of MS stars.
This suggests that to a given age spread should
correspond  an analogous gravity spread.

In \citet{dami14} it has been shown that for intermediate and low mass stars,
the $\gamma$ index depends on the effective temperature  in the sense that it
decreases at lower temperatures. This implies that
the  sensitivity of the  $\gamma$ index 
to the stellar gravity (and therefore to  age) 
increases towards later spectral types. 

For understanding if the observed age spread in NGC\,6530 corresponds to a gravity
 spread,
we compared the  $\gamma$ index of cluster members
with the $\gamma$ index typical of MS stars.
%Since the  distribution of stellar ages in NGC\,6530 increases
%rapidly toward younger ages, 
%we cannot adopt equal age ranges. 
%Alternatively,  
We performed the comparison by considering
CTTSe and WTTSp in 
two different age ranges by splitting the  objects 
 with ages younger and older than 1\,Myr, as shown in Fig.\,\ref{gammaages}.
 We note that we considered only the unbiased sample of members 
 with rotational velocity vsini$<50$\,km/s, 
$\gamma<1.01$ and 3800$<$T$_{\rm eff}/[K]<$5000. 

To avoid possible covariance effects 
due to the dependence on effective temperatures of both the  $\gamma$ index 
and  stellar ages, we performed this test using the 
stellar ages  obtained from 
the observed V and I magnitudes, corrected using the median reddening  
E(V-I)=0.5, % see output ages_pisa_phot
derived from Fig.\,\ref{intrinsiccolorspisa},
rather than the individual stellar reddenings,  that strongly depend on spectroscopic temperatures.
In this way, stellar ages  depend only on photometry 
while the $\gamma$ indices depend only on spectroscopy and  thus the two sets of parameters
are fully  statistically independent. This choice implies the usage of less
 certain stellar ages
but ensures the robustness  of the results. 
%In fact, random errors
%on the spectroscopic effective temperatures could led to estimate uncertain  stellar ages
%that could introduce an artifical  $\Delta \gamma$ and thus a false correlation.

%Even though
%the $\gamma$ indices show a quite large spread likely due to large errors affecting both
%  $\gamma$  and $T_{\rm eff}$, the  $\gamma$ values in each sample 
%show a decreasing trend
%towards lower effective temperatures, while lying above the locus of MS stars 
%$\gamma_{\rm MS}({\rm T}_{\rm eff})$.  

For each sample, we performed a second order
polynomial fit using the ROBUST\_POLY\_FIT idl routine
  and compared it with the MS reference locus
 $\gamma_{MS}(T_{\rm eff})$,
obtained using the $\gamma$$_{\rm MS} (\tau$)
and the $\tau$ index vs. T$_{\rm eff}$ calibration 
given in \citet{dami14} for MS stars.

For each sample, we computed the $\Delta \gamma$, defined as the difference
 between the best fit obtained
from our data and the reference locus of the MS stars, as shown in  
Fig.\,\ref{gammaages2}
as a function of T$_{\rm eff}$. In spite of the large uncertainties
affecting the best fit derivation,  the resulting $\Delta \gamma $
of the youngest CTTSe  is significantly larger than the one found for
 older CTTSe.  
Such difference is not found for WTTSp. All WTTSp, independently of their age,
  show a $\Delta \gamma$
consistent with the oldest CTTSe.

To test if the youngest and the oldest CTTSe 
(116 and 48 objects, % gamma_ages_ctt.pro
respectively) are really two different
populations, we performed a KS-test and found a
 probability of 3.5e-5 % gamma_ages_ctts.pro
 that they are taken from the same parent population.
  In contrast, the
 probability that the youngest and the oldest WTTSp 
 (44 and 38 objects, respectively) % gamma_ages_wtts.pro
 are two different
 populations is 0.77, % gamma_ages_wtts.pro
 suggesting that the latter belong to the same population.

The result on CTTSe (the majority of our sample) supports the previous conclusion that 
the observed age spread is  real and it
correlates with the gravity spread, as expected.  

We note that the observed gravity spread
is obtained also assuming stellar ages derived using  
  spectroscopic 
effective temperatures but, for the reasons discussed above, 
it is less robust than that shown in Fig.\,\ref{gammaages}.

\section{Discussion}
Figures\,\ref{memcliiiagebinmap} and \ref{memcliiagebinmap} show the
spatial distribution of cluster members classified, respectively,
as WTTSp   and CTTSe,
split into the two age ranges used in the previous section and
 as a function of the proper motion in right
 ascension $\mu_\alpha$. 
 
These  distributions indicate that 
the  WTTSp are sparsely distributed  but few of them fall in the region
where the youngest CTTSe are found, i.e. around 9\,Sgr and Her\,36,
and along the S-E bright rim to the south of the cluster center.
Even the CTTSe with age older than 1\,Myr  
do not show strong evidence of clustering  while
the CTTSe formed in the last 1\,Myr  
show a  pattern with  two radial concentrations.
The  most populated one is located
 around the compact core of the cluster NGC\,6530, 
 while the second group is
 in the 9Sgr/Her36  region. In addition, there is a group of CTTSe  members 
formed in the last 1\,Myr, that follows the bow-shape
 structure connecting the Her\,36 region with 
M8E-IR and HD165052 through the bright  H$\alpha$ rim.
The CTTSe of the latter group and those found around 9\,Sgr
have proper motions $\mu_\alpha$
slightly larger than those of the other members, in agreement with
\citet{dami18a} and  Wright et al. (2019 ,submitted).
This is an evidence that these latter objects move with
a slightly  different motion  with respect to the other members.

Our findings on the spatial distribution and previous results
 are used here to evaluate two different scenarios.
The first scenario is  based on the assumption that
our stellar age uncertainties are underestimated and that 
there is not age spread,  
i.e.  all members  formed in the same SF event.    
    According to  this scenario, the $\gamma$ spread found in CTTSe
    could still be interpreted  as a spread in 
    gravity and hence in radius. As is known, low gravity stars in the PMS phase
    appear more luminous than other stars of the same mass since,
   as predicted by theoretical models, they descend almost vertically
    with time along the tracks.
   These low-gravity objects, for some intrinsic reason
    (episodic accretion, 
   magnetic fields, starspots),
   descend along the tracks more slowly with respect
   to high-gravity stars of the same age and therefore we 
   underestimate their ages,
   and we mis-interpret a luminosity spread as an age spread \citep[e.g.][]{bara15}.
   However, we observe that
    these objects show a very peculiar spatial pattern with
    respect to the high-gravity
   CTTSe stars that, in contrast, are sparsely distributed.
    In addition, the subsample of
   stars along the S-E bright rim and those around 9\,Sgr
   can be  kinematically distinguished from
    the stars concentrated around the cluster center.
    If we assume that the gravity-spread is due
    to an intrinsic property of the star (episodic accretion,
    magnetic fields or  starspots), there is not any reason
 for which  it  should be connected to
  the spatial regions and/or to the kinematics of these stars.
   
An alternative more convincing scenario is that 
 that SF occured   within few Myrs but not within a single 
burst.
  In this scenario, 
the gamma spread in the HR diagram is evidence of gravity 
    (and radius) spread due to  different stellar contraction stages,
    i.e. due to an intrinsic age spread.
  The youth of these members is supported by their 
  spatial distribution   
   around the cluster center,  where several authors have suggested ongoing 
   star formation. In fact, this region corresponds to the CO bright spot found 
   by \citet{lada76} that is receding from us
           with an heliocentric RV ($\sim$6km/s). 
The CO gas  moves away from the cluster center  since its RV
            is significantly redshifted with respect to
	     the cluster RV$_{\rm cl}$=0.17 km/s. 
           This suggests that very recent SF events are expected there due 
	   to enhanced  compression of the gas.
       
The second group of youngest members is distributed 
along the bright rim to the south of the cluster center where CO and dust maps
    clearly show the presence of a dense molecular cloud. The arch along which 
    the youngest
    stars are distributed starts from the binary star
    HD 165052  \citep[O6.5 V + O7.5 V][]{aria02}  and ends near the south edge of the
    Great Rift, i.e. the dark lane that splits the optical nebula.
 Finally, the remaining youngest stars are found 
 around 9\,Sgr, the ionizing star in the Hourglass Nebula, the compact and dense
   \hii region,  powered  also by the O7 star Herschel 36.
   
A further clue supporting the latter scenario is 
the different kinematics  of the stars in S-E bright rim and  around 9\,Sgr 
with respect to that of stars in
      the cluster center that can be interpreted as evidence
       of two different SF events, pushed and likely
      triggered by two different ionizing fronts, giving origin in the last 1\,Myr 
      to  different morphology 
      subclusterings, one circularly concentrated (in the cluster center)
      and one elongated, following
      the filamentary morphology of the CO submillimetre arch. 
      This scenario has been already invoked
      in the past in the literature \citep[see][for a review]{toth08}.   

Our conclusion confirms  the \citet{dami14} finding that the $\gamma$ index
%and thus the stellar gravity, 
undergoes  sensitive changes in the first Myr, 
for low mass stars with T$_{\rm eff}\lesssim$5000\,K. This is in agreement
 with the expected increase of  gravity during the PMS contraction. 
%The evidence of a gravity spread supports the hypothesis of a real age spread 
%in NGC\,6530 rather than
%an apparent spread due to phometric or spectroscopic uncertainties.

\section{Summary and conclusions}
Spectroscopic GES parameters cross-correlated with literature data allowed 
us to identify \memconf confirmed members.
These include \memctts sources with
signatures of accretion and/or circumstellar disks.
For the sample of non-members, we used
 the gravity-sensitive $\gamma$
 index to distinguish the populations of dwarfs and giants. 
 The derived interstellar reddening obtained from spectroscopic 
effective temperatures,
fit in a coherent picture with the inferred luminosity class of the
observed objects. In fact,  most of the giant stars are also the more
reddened ones, since
they are mainly located behind the nebula, as expected.
 The foreground stars are mainly MS stars with spectral types earlier 
 than those of giants.
 \emph{Gaia} DR2 parallaxes confirm that giants are located behind 
the nebula while late type (early)  MS stars are  mainly 
located in front of (behind) the nebula.

In agreement with previous results from literature, the mean reddening affecting
the NGC\,6530 cluster members is relatively low and this confirms that the cluster
is in front of the surrounding nebula. There are, however, a few members affected by large
reddening, that, in contrast, are  embedded in the dust of the Lagoon Nebula.

Photometric diagrams and spectroscopic reddening ratios in different bands
allowed us to confirm in two independent ways, that the reddening law suitable
for cluster members is the non-standard R$_V$=5.0. The result holds
only for cluster members and it is likely driven by the complex dust structures
around the cluster members. %This is coherent with the indication
The analogous analysis performed
for the background giants, located behing the nebula, shows that these,
 in contrast,
obey to a law more similar to the standard reddening law. 

The spatial distribution of cluster members and of contaminant MS and giant stars
is a clear signature of the three-dimensional structure of the nebula, with 
evidence for darker areas where no background stars are detected and of
more transparent  zones, where  reddened background giants 
are found. Cluster members do not show a regular radial spatial distribution,
but form a main compact  peak   surrounded by several 
secondary peaks, mainly in the Her\,36 region and around the S-E bright rim.

Stellar ages inferred from our data suggest that cluster members,
both WTTSp and CTTSe, are characterized by
similar age distributions. This supports that measured stellar ages are not
affected by non-photospheric 
effects due to the presence of  circumstellar disks and/or accretion.
 The age distribution suggests that star formation started about 15\,Myr ago
 but formed the bulk of the stars  in the last 1-2\,Myr.

For low mass stars, 
observational uncertainties on isochronal ages are significantly smaller than
the observed age dispersion and thus this was 
interpreted as a real age spread.
This result has been confirmed by the evidence
of a similar spread in the gravity-sensitive $\gamma$ index
 and by the star spatial distribution and kinematics.

As already found by \citet{kala15}, CTTSe and WTTSp members show a quite
 different spatial distribution.
However, while stars older than 1-2\,Myr might be found away from
where they were formed,
 objects formed in the last 1\,Myr are likely found in the same place where
  they likely formed.
  Instead, 
 assuming that all stars are observed where they formed,
our results would confirm 
 previous studies suggesting a scenario of  sequential star formation.
In fact, the oldest stars of our sample,
 are sparsely distributed around the cluster center while stars in the Her\,36
region and around the S-E bright rim appeared only in the last 1\,Myr.

This result would be in contrast with the conclusions derived in \citet{kala15}
who do not find evidence supporting the sequential star formation.
We note, however, that their results are based on a sample of members including
only accretors, while our sample includes both CTTSe and WTTSp  members.
The lack of a large fraction of these latter objects, mainly distributed 
in region where the nebular H$\alpha$ emission is lower, might explain 
their conclusion.

%We note, however, that the scenario of  sequential star formation 
%cannot be  considered conclusive but could be
% tested by a further dynamical study  (Wright et al. in preparation).
The very peculiar spatial pattern formed by the
 very young CTTSe  members, that is clearly
correlated to the position of
 the O-type stars, can be interpreted as evidence of  triggered
star formation by the O-star ionization fronts.
In this context, it may be useful to recall that the spatial distribution of
 the youngest stars is nearly coincident with that of two expanding 
 shells of ionized gas \citep{dami17a}, likely driven by the massive 
 stars in the cluster core  and the protostar M8E-IR, respectively. M8-IR is an 
 embedded massive protostar, much younger than cluster core stars,
 and its shell (of which only the receding portion was detected, 
 while the approaching part is blocked by the dust in the SE bright rim)
 is probably younger than that around cluster core stars. 
 A probable combined effect of the two oppositely-directed shells
 is an enhanced compression of the local diffuse medium.
 The near-coincidence of this region of enhanced compression 
 with the overdensity of stars younger than 1\,Myr seen in 
 Fig.\,\ref{memcliiagebinmap}, but not with cluster stars 
 in older age ranges, is highly suggestive that such compression 
 (ultimately driven by the massive stars' winds) might have been 
 the dominant cause of the most recent star-formation episode in 
 this part of the Nebula.
 
\begin{acknowledgements}
 We wish to thank the anonymous referee for the helpful suggestions. 
This research has made use of data  products from observations made with ESO
 Telescopes at the La Silla Paranal Observatory under Programme ID 188.B-3002.
  These data products have been processed by the Cambridge Astronomy Survey 
  Unit (CASU) at the Institute of Astronomy, University of Cambridge, and by
   the FLAMES/UVES reduction team at INAF/Osservatorio Astrofisico di Arcetri.
    These data have been obtained from the Gaia-ESO Survey Data Archive, 
    prepared and hosted by the Wide Field Astronomy Unit, Institute for 
    Astronomy, University of Edinburgh, which is funded by the UK Science 
    and Technology Facilities Council.
 The authors acknowledge support through the PRIN INAF 2014 funding scheme
of the National Institute for Astrophysics (INAF) of the Italian Ministry
 of Education, University and Research ("The \emph{Gaia}-ESO Survey",
  P.I.: S. Randich). 
This work was partly supported by the European Union FP7 programme through
 ERC grant number 320360 and by the Leverhulme Trust through grant RPG-2012-541.
  We acknowledge the support from INAF and Ministero dell'Istruzione, 
  dell'Universit\`a e della Ricerca (MIUR) in the form of the grant
   "Premiale VLT 2012". 
   E.T. acknowledges University of Pisa ("Modelli di stelle di massa 
piccola-intermedia per la determinazione dell'et\`a degli 
ammassi stellari osservati dal satellite \emph{Gaia}" 
PI: S. Degl'Innocenti, 2018) and INFN ("Iniziativa 
specifica TAsP").
   The results presented here benefit from discussions
    held during the Gaia-ESO workshops and conferences supported by the ESF 
    (European Science Foundation) through the GREAT Research Network Programme.
This work has made use of data from the European Space Agency (ESA) mission
\emph{Gaia} (\url{https://www.cosmos.esa.int/gaia}), 
processed by the \emph{Gaia}
Data Processing and Analysis Consortium (DPAC,
\url{https://www.cosmos.esa.int/web/gaia/dpac/consortium}). 
Funding for the DPAC
has been provided by national institutions, in particular the institutions
participating in the \emph{Gaia} Multilateral Agreement.
The VPHAS+ mosaics were generated using the
MONTAGE software maintained by NASA/IPAC.
\end{acknowledgements}
\bibliographystyle{aa}
\bibliography{/home/prisinzano/BIBLIOGRAPHY/bibdesk}
 \begin{landscape}
\begin{table}
\caption{Literature data and Gaia-ESO survey parameters of cluster members.
The full table is available in electronic format.
 \label{tablemem1}}
\centering
\begin{tabular} {c c c c c c c c c c c}
\hline\hline
CNAME & ID-VPHAS & RA  & Dec & V   & I   & RV   & vsini& EW(Li) & FWZI  & X det. \\
      &          & deg & deg & mag & mag & km/s & km/s & m$\AA$ & $\AA$ &        \\
\hline
18035527-2417000 & 18035527-2417000 &    270.98012 &    -24.28331 &   14.700$\pm$   0.002 &  13.429$\pm$   0.012 &   -5.22 $\pm$    0.20 &   66.18 $\pm$    6.62 &  319.30 $\pm$    5.55 &    3.22 $\pm$    1.63 &  1 \\
18040419-2428321 & 18040418-2428321 &    271.01746 &    -24.47558 &   13.745$\pm$   0.021 &  12.844$\pm$   0.010 &    3.79 $\pm$    0.46 &   18.72 $\pm$    1.87 &   15.70 $\pm$    2.90 &    8.87 $\pm$    1.50 &  0 \\
18040693-2420251 & 18040692-2420253 &    271.02887 &    -24.34031 &   18.634$\pm$   0.011 &  16.261$\pm$   0.006 &   -3.25 $\pm$    0.85 &   18.83 $\pm$    1.88 &  502.05 $\pm$   16.33 &   13.78 $\pm$    1.49 &  0 \\
18041245-2411516 & 18041245-2411515 &    271.05187 &    -24.19767 &   17.301$\pm$   0.011 &  15.465$\pm$   0.005 &   -1.50 $\pm$    1.26 &   20.26 $\pm$    2.03 &  349.94 $\pm$   80.11 &   15.52 $\pm$    2.78 &  0 \\
18041306-2423173 & 18041306-2423174 &    271.05433 &    -24.38806 &   17.471$\pm$   0.003 &  15.547$\pm$   0.004 &    3.34 $\pm$    0.22 &   20.32 $\pm$    2.03 &  420.20 $\pm$    5.39 &   15.25 $\pm$    2.77 &  1 \\
18041479-2421056 & 18041481-2421059 &    271.06162 &    -24.35156 &   17.902$\pm$   0.008 &  15.656$\pm$   0.003 &    4.33 $\pm$    0.21 &   17.62 $\pm$    1.76 &  553.10 $\pm$   11.29 &    5.18 $\pm$    0.47 &  1 \\
18041907-2427590 & 18041907-2427590 &    271.07946 &    -24.46639 &   18.463$\pm$   0.012 &  15.676$\pm$   0.002 &    3.40 $\pm$    1.55 &   15.54 $\pm$    1.55 &  350.65 $\pm$   14.45 &   12.08 $\pm$    1.27 &  0 \\
18042497-2416013 & 18042497-2416013 &    271.10404 &    -24.26703 &   18.614$\pm$   0.009 &  16.501$\pm$   0.008 &   -2.78 $\pm$    0.74 &   25.83 $\pm$    2.58 &  460.40 $\pm$   15.84 &   18.57 $\pm$    1.66 &  0 \\
18042527-2422510 & 18042528-2422510 &    271.10529 &    -24.38083 &   16.417$\pm$   0.005 &  14.501$\pm$   0.003 &    0.67 $\pm$    0.27 &   17.82 $\pm$    1.78 &  508.90 $\pm$    5.94 &    9.77 $\pm$    1.14 &  1 \\
18042552-2420390 & 18042551-2420391 &    271.10621 &    -24.34408 &   16.819$\pm$   0.044 &  14.953$\pm$   0.004 &    0.28 $\pm$    0.22 &   21.06 $\pm$    2.11 &  417.20 $\pm$    8.06 &   16.78 $\pm$    1.11 &  1 \\
18042914-2421526 & 18042914-2421527 &    271.12133 &    -24.36456 &   14.943$\pm$   0.004 &  13.479$\pm$   0.003 &   -1.34 $\pm$    2.12 &   45.97 $\pm$    4.60 &  361.10 $\pm$    5.50 &    6.27 $\pm$    1.02 &  1 \\
18043021-2417475 & 18043020-2417473 &    271.12587 &    -24.29653 &   18.365$\pm$   0.005 &  16.172$\pm$   0.004 &   -0.48 $\pm$    0.66 &   18.06 $\pm$    1.81 &  579.90 $\pm$   14.20 &    7.55 $\pm$    2.47 &  1 \\
18043043-2428534 &  -- &    271.12679 &    -24.48150 &   18.289$\pm$   0.018 &  15.930$\pm$   0.004 &    6.57 $\pm$    1.51 &   43.20 $\pm$    4.32 &  716.65 $\pm$  103.55 &   13.58 $\pm$    5.16 &  1 \\
18043172-2414540 & 18043174-2414543 &    271.13217 &    -24.24833 &   18.121$\pm$   0.005 &  15.597$\pm$   0.005 &   22.92 $\pm$    7.55 &   75.48 $\pm$    7.55 &  687.09 $\pm$  201.00 &    6.38 $\pm$    1.09 &  1 \\
18044089-2417106 & 18044088-2417108 &    271.17025 &    -24.28628 &   16.740$\pm$   0.003 &  14.732$\pm$   0.002 &   -3.36 $\pm$    1.33 &   22.55 $\pm$    2.25 &  448.10 $\pm$    5.60 &   12.22 $\pm$    5.81 &  1 \\
18044363-2427591 &  -- &    271.18179 &    -24.46642 &   18.509$\pm$   0.009 &  15.949$\pm$   0.006 &   -5.96 $\pm$    2.63 &   28.21 $\pm$    2.82 &  475.50 $\pm$   17.45 &   16.53 $\pm$    3.86 &  0 \\
18044404-2419390 &  -- &    271.18350 &    -24.32750 &   16.297$\pm$   0.003 &  14.391$\pm$   0.006 &    0.37 $\pm$    0.65 &   20.19 $\pm$    2.02 &  517.00 $\pm$    6.93 &    7.42 $\pm$    0.19 &  1 \\
18044417-2415250 &  -- &    271.18392 &    -24.25692 &   17.151$\pm$   0.004 &  14.802$\pm$   0.005 &    9.32 $\pm$    0.28 &   50.66 $\pm$    5.07 &  527.60 $\pm$   11.03 &   11.42 $\pm$    0.16 &  1 \\
18044497-2413110 & 18044500-2413111 &    271.18737 &    -24.21972 &   18.719$\pm$   0.038 &  16.159$\pm$   0.025 &   -2.63 $\pm$    1.24 &   22.94 $\pm$    2.29 &  439.40 $\pm$    8.48 &    9.00 $\pm$    1.30 &  1 \\
18044524-2420585 & 18044524-2420585 &    271.18833 &    -24.34958 &   16.185$\pm$   0.003 &  14.352$\pm$   0.008 &   -1.68 $\pm$    1.17 &   22.43 $\pm$    2.24 &  508.80 $\pm$    6.58 &    6.63 $\pm$    1.59 &  1 \\
... & ... & ... & ...  & ... & ...   & ...  & ...  & ... & ... & ... \\ 
\hline
\end{tabular}
\tablefoot{ Col.\,1: Cname, col.\,2: identification in the VPHAS catalog,
col.s\,3 and 4: equatorial coordinates, col.\,5 and 6: WFI photometry 
from \citet{pris05}, col.\,7 and 8: radial and rotational velocities,
col.\,9 and 10 equivalent width of the lithium line and FWZI of the H$\alpha$ line, 
col.\,11 flag of X-ray detection (1=detection, 0=no detection, -1=undefined)}
\end{table}
\end{landscape}

 \begin{landscape}
\begin{table}
\caption{Gaia-ESO survey and Gaia parameters, individual reddenings and isochronal
 ages of cluster members. The full table is available in electronic format.
 \label{tablemem2}}
\centering
\begin{tabular} {c c c c c c c c c c}  
\hline\hline
CNAME & $\gamma$ & T$_{\rm eff}$ & $\mu_{\alpha}$  & $\mu_{\delta}$ &
$\pi$ & E(V-I) & Log(t) & Mem. & Class\\
      &          & K             &   mas/yr         & mas/yr        &
 mas & mag     & yr     &   &  \\
\hline
18035527-2417000 &     1.00 $\pm$    0.00 & 4992.00 $\pm$   76.00 &   1.093 $\pm$   0.060 &  -1.897 $\pm$   0.049 &   0.724 $\pm$   0.034 &    0.27 $\pm$    0.04 &    5.95 $\pm$    0.11 &M &WTTSp \\
18040419-2428321 &     1.00 $\pm$    0.00 & 5117.00 $\pm$   41.84 &   2.428 $\pm$   0.053 &  -6.415 $\pm$   0.042 &   0.943 $\pm$   0.032 &   -0.04 $\pm$    0.03 &    6.08 $\pm$    0.09 &M &CTTSe \\
18040693-2420251 &     0.89 $\pm$    0.02 & 3931.00 $\pm$   50.36 &   1.669 $\pm$   0.277 &  -2.970 $\pm$   0.229 &   0.463 $\pm$   0.189 &    0.59 $\pm$    0.05 &    6.10 $\pm$    0.08 &M &CTTSe \\
18041245-2411516 &     0.94 $\pm$    0.01 & 3915.00 $\pm$  107.29 &   0.745 $\pm$   0.137 &  -1.691 $\pm$   0.118 &   0.860 $\pm$   0.079 &    0.04 $\pm$    0.11 &    6.19 $\pm$    0.09 &M &CTTSe \\
18041306-2423173 &     0.96 $\pm$    0.01 & 4128.00 $\pm$  181.00 &   1.186 $\pm$   0.173 &  -2.777 $\pm$   0.146 &   0.724 $\pm$   0.098 &    0.33 $\pm$    0.16 &    6.16 $\pm$    0.10 &M &CTTSe \\
18041479-2421056 &     0.91 $\pm$    0.01 & 4023.00 $\pm$   48.67 &   1.064 $\pm$   0.193 &  -1.879 $\pm$   0.159 &   0.632 $\pm$   0.138 &    0.56 $\pm$    0.05 &    5.86 $\pm$    0.08 &M &CTTSe \\
18041907-2427590 &     0.88 $\pm$    0.01 & 3162.00 $\pm$   23.81 &   1.290 $\pm$   0.216 &  -2.289 $\pm$   0.173 &   0.812 $\pm$   0.128 &    0.04 $\pm$    0.04 &    5.28 $\pm$    0.13 &M &CTTSe \\
18042497-2416013 &     0.99 $\pm$    0.01 & 4232.00 $\pm$   62.54 &   2.850 $\pm$   0.319 &  -2.175 $\pm$   0.259 &   0.806 $\pm$   0.153 &    0.60 $\pm$    0.06 &    6.73 $\pm$    0.08 &M &CTTSe \\
18042527-2422510 &     0.99 $\pm$    0.00 & 4253.00 $\pm$   17.00 &   0.931 $\pm$   0.095 &  -2.378 $\pm$   0.076 &   0.749 $\pm$   0.050 &    0.42 $\pm$    0.02 &    5.57 $\pm$    0.06 &M &CTTSe \\
18042552-2420390 &     0.98 $\pm$    0.01 & 4252.00 $\pm$   16.00 &   0.922 $\pm$   0.148 &  -2.169 $\pm$   0.127 &   0.575 $\pm$   0.086 &    0.37 $\pm$    0.05 &    5.85 $\pm$    0.10 &M &CTTSe \\
18042914-2421526 &     0.99 $\pm$    0.00 & 4992.00 $\pm$    6.00 &   0.902 $\pm$   0.073 &  -1.346 $\pm$   0.060 &   0.721 $\pm$   0.042 &    0.46 $\pm$    0.01 &    5.80 $\pm$    0.09 &M &CTTSe \\
18043021-2417475 &     1.00 $\pm$    0.02 & 4018.00 $\pm$   61.17 &   0.748 $\pm$   0.198 &  -1.915 $\pm$   0.160 &   0.684 $\pm$   0.114 &    0.50 $\pm$    0.06 &    6.27 $\pm$    0.09 &M &CTTSe \\
18043043-2428534 &     0.92 $\pm$    0.01 & 3848.00 $\pm$   39.90 &   1.563 $\pm$   0.265 &  -2.135 $\pm$   0.208 &   0.430 $\pm$   0.136 &    0.50 $\pm$    0.05 &    5.91 $\pm$    0.07 &M &CTTSe \\
18043172-2414540 &     0.93 $\pm$    0.01 & 3691.00 $\pm$   19.53 &   0.848 $\pm$   0.184 &  -1.927 $\pm$   0.157 &   0.978 $\pm$   0.099 &    0.48 $\pm$    0.02 &    5.67 $\pm$    0.06 &M &CTTSe \\
18044089-2417106 &     1.00 $\pm$    0.01 & 4246.00 $\pm$   31.00 &   1.122 $\pm$   0.187 &  -1.577 $\pm$   0.171 &   1.202 $\pm$   0.099 &    0.51 $\pm$    0.03 &    5.59 $\pm$    0.07 &M &CTTSe \\
18044363-2427591 &     0.86 $\pm$    0.02 & 3795.00 $\pm$  103.68 &   2.726 $\pm$   0.223 &  -1.951 $\pm$   0.193 &   0.869 $\pm$   0.109 &    0.64 $\pm$    0.12 &    5.76 $\pm$    0.06 &M &CTTSe \\
18044404-2419390 &     0.99 $\pm$    0.01 & 4395.00 $\pm$   19.00 &   1.380 $\pm$   0.095 &  -1.736 $\pm$   0.077 &   0.765 $\pm$   0.049 &    0.54 $\pm$    0.02 &    5.55 $\pm$    0.06 &M &CTTSe \\
18044417-2415250 &     0.98 $\pm$    0.01 & 3791.00 $\pm$   13.79 &   2.124 $\pm$   0.121 &  -0.932 $\pm$   0.102 &   0.792 $\pm$   0.069 &    0.43 $\pm$    0.02 &    5.03 $\pm$    0.14 &M &CTTSe \\
18044497-2413110 &     0.98 $\pm$    0.01 & 4216.00 $\pm$   48.00 &   2.763 $\pm$   0.331 &  -0.595 $\pm$   0.276 &   0.812 $\pm$   0.158 &    1.04 $\pm$    0.06 &    5.85 $\pm$    0.10 &M &CTTSe \\
18044524-2420585 &     1.00 $\pm$    0.00 & 4312.00 $\pm$   95.00 &   1.480 $\pm$   0.129 &  -1.938 $\pm$   0.103 &   0.719 $\pm$   0.052 &    0.39 $\pm$    0.08 &    5.57 $\pm$    0.06 &M &CTTSe \\
... & ... & ... & ... & ... & ... & ... & ... & ... & ... \\
\hline
\end{tabular}
\tablefoot{ Col.\,1: Cname, col.\,2: gravity-sensitive $\gamma$  index,
col.\,3 effective temperatures, col.\,4, 5, and 6: Gaia 
proper motions and parallaxes, col.s\,7 and 8: individual reddenings and 
isochronal ages derived in this work, col.\,9: member type (M=confirmed member, PM=probable member),
col.\,10: WTTSp or CTTSe classification.}
\end{table}
\end{landscape}

\end{document}